\documentclass[12pt,preprint]{aastex}

\shorttitle{SGMCs in Antennae}
\shortauthors{Wilson et al.}

\begin{document}
\newcommand{\pad}{\phantom{$0$}}

\title{The Mass Function of Super Giant Molecular Complexes and Implications
for Forming Young Massive Star Clusters in the Antennae (NGC 4038/39)}
\author{Christine D. Wilson\altaffilmark{1},
Nicholas Scoville\altaffilmark{2},
Suzanne C. Madden\altaffilmark{3},
Vassilis Charmandaris\altaffilmark{4,5}}

\altaffiltext{1}{Department of Physics and Astronomy, McMaster University,
Hamilton, Ontario L8S 4M1 Canada} 
\altaffiltext{2}{Division of Physics, Mathematics, and Astronomy,
Caltech 105-24, Pasadena CA 91125 U.S.A.}
\altaffiltext{3}{CEA/DSM/DAPNIA/Service d'Astrophysique, 
CE-Saclay, 91191 Gif sur Yvette Cedex, France}
\altaffiltext{4}{Astronomy Department, Cornell University, Ithaca NY 14853 
U.S.A.}
\altaffiltext{4}{Chercheur Associ\'e, Observatoire de Paris, LERMA, 61 Av. de
l'Observatoire, F-75014 Paris, France}

\email{wilson@physics.mcmaster.ca}

% August 27, 2003: downloaded submited version from web to make final changes

\begin{abstract}

We have used previously published observations of the CO emission
from the Antennae (NGC 4038/39) to study the detailed properties
of the super giant molecular complexes with the goal of understanding
the formation of young massive star clusters. Over
a mass range from $5\times 10^6$ to $9\times 10^8$ M$_\odot$, the molecular
complexes follow a power-law mass function with a slope of
$-1.4 \pm 0.1$, which is very similar to the slope seen at lower masses
in molecular clouds and cloud cores in the Galaxy. Compared
to the spiral galaxy M51, which has a similar surface density and
total mass of molecular gas, the Antennae contain clouds that
are an order of magnitude more massive. 
%**vv
%One possibility is that
%the formation of very massive gas clouds in M51 is suppressed
%by high shear in its smoothly rotating disk. 
%**^^
Many of the youngest
star clusters lie in the gas-rich overlap region, where
extinctions as high as $A_v \sim 100$ mag imply that the
clusters must lie in front of the gas. Young clusters
found in other regions of the galaxies can be as far as 2 kpc from
the nearest massive cloud, which suggests that either young clusters
can form occasionally in clouds less massive than $5\times 10^6$ M$_\odot$
or these young clusters have already destroyed their parent
molecular clouds. Combining data on the young clusters, thermal
and nonthermal radio sources, and the molecular gas suggests
that young massive clusters could have formed at a constant rate
in the Antennae over the last 160 Myr and that sufficient gas 
exists to sustain this cluster
formation rate well into the future. However, this conclusion
requires that a very high fraction of the massive
clusters that form initially in the Antennae do not survive as long as
100 Myr. Furthermore, if most young massive clusters do
survive for long periods,  the Antennae must be experiencing
a relatively short burst of cluster formation to prevent
the final merger remnant from exceeding the observed specific
frequency of star clusters in elliptical galaxies by a wide margin.
Finally, we compare our data with two models for massive star
cluster formation and conclude that the model where young massive
star clusters form from dense cores within the observed
super giant molecular complexes is most consistent with our
current understanding of this merging system.

\end{abstract}
\keywords{galaxies: individual (NGC 4038, NGC 4039, Arp 244) --- galaxies: ISM --- ISM: molecules --- stars: formation --- radio lines: galaxies}

\section{Introduction}

The discovery of very luminous young star clusters in merger remnant
galaxies \citep{h92,w93,ws95}
has led to a dramatic shift in our understanding of star
cluster formation. 
Some of these young star clusters may be massive
enough \citep{zf99,hf96,m02} that they can be 
viewed as young counterparts to the ubiquitous
globular clusters, which are found in nearly every extragalactic system
from giant cD elliptical galaxies to dwarf irregular galaxies 
\citep{harris01}. The
discovery of these young massive 
clusters suggests that massive star cluster formation
was not a process confined exclusively to the early universe
(e.g. \citet{c01,fr85}), but
rather one that continues on to the present day. Thus, while 
attempting to understand the 
formation and evolution of young massive star clusters is interesting
in its own right, such studies may also shed light on the process
of globular cluster formation, which clearly occurred in a wide variety
of galactic environments in the early universe.

Since star clusters in the Milky Way are observed to form in dense clouds
of molecular hydrogen gas, it is likely that these young massive star clusters
formed in a similar way. However, the relatively large masses
($> 10^5$ M$_\odot$) estimated for these clusters  are comparable
to the masses of typical giant molecular clouds in nearby galaxies
\citep{sss85}.
Since individual molecular clouds typically form stars with efficiencies
of only a few percent \citep{el91}, 
these large cluster masses pose a challenge to our current 
understanding of star formation. In particular, if the star clusters
are gravitationally bound, the star formation efficiency in the 
material from which they formed must have been closer to 50\%.
Such high star formation efficiency is typically observed only in
smaller bound cores within individual giant molecular clouds
\citep{l91a}. These
facts have led to two quite different explanations for how these
young massive star clusters formed. To explain the formation
of globular clusters, \citet{hp94} postulated the
existence of ``super giant molecular clouds'' in the early universe.
With masses up to $10^9$ M$_\odot$, but internal structures similar
to local giant molecular clouds, these massive clouds could contain
correspondingly massive cores, which could form massive (bound) star
clusters. If similarly massive clouds exist in galaxies in the local universe,
this might also be a viable method for forming young massive star clusters.
An alternative explanation for young massive star clusters proposed
by \citet{s96} involves the collapse of a pre-existing
giant molecular cloud to form a single massive star cluster. Such
collapse could be triggered by overpressure in the diffuse
interstellar medium produced during the
collision of two massive galaxies \citep{js92}, but
would require a collapse of 2-3 orders of magnitude in linear size. Clearly,
one way to shed some light on the processes which lead to the
formation of young massive
star clusters is to study the properties of the molecular
gas in galaxies in which such clusters are found.

Since the original discoveries of rich populations of young massive star
clusters in nearby merging and merger-remnant galaxies, 
young massive 
clusters have been identified in a very wide range of galactic
environments, from dwarf starburst galaxies (i.e. M82, \citet{o95}) to
spiral galaxies (i.e. M51, \citet{l00}) to merging galaxies in all
stages of evolution (see \citet{w01} for a complete list).
The Antennae system (NGC 4038/39, Arp 244) 
has the largest number of young massive star clusters identified
in a single galaxy to date \citep{ws95,w99}. In addition, the 
clusters in the Antennae are spread over a large area of
the galaxy ($\sim 100$ kpc$^2$), which implies that these clusters have
formed over much of the galactic disks. The Antennae
have a higher surface density of young massive star clusters than older merger
systems, although smaller galaxies such as M82 \citep{o95}
have a larger absolute surface density of clusters confined
to a relatively small region. This extremely
rich population of clusters, combined with the proximity of
the Antennae system (19 Mpc for $H_o = 75$ km s$^{-1}$ Mpc$^{-1}$),
makes it the logical first choice for detailed studies of
the structure and properties of the molecular interstellar
medium with the aim of understanding the formation of young massive star
clusters.

The first detection of CO in the Antennae by \citet{sm85} suggested
that the system had a relatively high ratio of far-infrared to CO
luminosity. More recent complete mapping of the two galactic disks
by \citet{g01} revealed a much larger reservoir of molecular gas,
which led to the suggestion that this system could possibly 
evolve into an ultraluminous starburst as the merging becomes more
advanced. Early interferometric CO observations of the Antennae system
were presented by \citet{s90}. These observations revealed that 
the most massive concentration of molecular gas lay outside the two
nuclei in a region which Stanford et al. named ``the overlap region''.
In a recent paper (Wilson et al. 2000, hereafter
\citet{w00}), we presented
interferometric CO observations of the Antennae with improved
resolution and sensitivity and covering a larger fraction of
the galactic disks. These data show the presence of massive
($> 10^8$ M$_\odot$), gravitationally bound gas clouds, which
we term super giant molecular complexes \citep{w00}. A detailed
comparison of the CO data with high-resolution mid-infrared
data from the {\it Infrared Astronomical Satellite} \citep{v96,
m98} suggests that the extremely bright mid-infrared
source seen in the overlap region could be produced by 
star formation induced by the collision
of two of these massive complexes (however, see \citet{l01}
for evidence that the clouds may not be colliding). An alternative
explanation is that this region contains extremely young ($<1$ Myr)
sites of star formation \citep{w00}.

In this paper, we use the data set presented in \citet{w00} for a detailed
study of the population of molecular complexes in the Antennae,
such as their mass function, their spatial overlap with the young massive
star clusters, and a comparison with theories of star cluster formation.
In \S\ref{obs} we describe the observations and data reduction,
as well as a brief discussion of how the clouds were identified from
the data cube. 
In \S\ref{mass} we present the cloud mass spectrum and
compare our results to observations of massive cloud complexes in
other nearby galaxies. We also discuss what can be
learned from the relative locations of the young massive star clusters and
the super giant molecular complexes. 
In \S\ref{M51comp}, we make a detailed comparison
of the Antennae with the complexes and star clusters observed in
M51 \citep{rk90,l00,s01}. In \S\ref{short-lived} we discuss the recent
star formation history in the Antennae and in \S\ref{SSCF} we compare the
properties of the molecular complexes with two models for massive
star cluster formation. The paper
is summarized in \S\ref{concl}.

\section{Analysis \label{obs}}

\subsection{Observations and Data Reduction}

The CO observations of the Antennae originally presented in \citet{w00}
were obtained with the Caltech Millimeter Array between 1998 March
and 1999 February. Three overlapping field centers were chosen to
cover the majority of the young massive 
star clusters identified by \citet{ws95}.
The coordinates of the three field centers were 
specified in 1950 coordinates as
(11:59:20.3, -18:35:31) (Field 1),
(11:59:20.3, -18:36:11) (Field 2) and (11:59:17.5, -18:35:21) (Field 3), 
which gives an
offset between adjacent field centers of 40$^{\prime\prime}$.
In J2000 coordinates, the equivalent field centers are
(12:01:54.09, -18:52:13.25), (12:01:54.09,  -18:52:53.25), and
(12:01:51.28,  -18:52:03.25).
Each field was observed for the equivalent of one complete transit in
each of four configurations: C (compact), L (low), E (equatorial) and
H (high resolution). The baselines for these observations ranged from
15 m to 250 m.
The observations of Fields 1 and 2 were interleaved
in each observing track, which results in very similar {\it uv} coverage
for these two fields. Field 3 (which was added later) was observed
in separate tracks. These CO data represent approximately 90 hours
observing time with the array.

The spectrometer was configured to give 2 MHz (5.2 km s$^{-1}$)
resolution and was tuned to a central velocity of 1547 km s$^{-1}$;
the resulting velocity range was from 1256 km s$^{-1}$ to 1838 km s$^{-1}$.
The system temperatures obtained during these observations range from
1000 to 2000 K (single sideband), with an average value of 1500 K.
(Note that these system temperatures are substantially higher than those
typically obtained by the array due to the low declination of the source.)
The data calibration was performed using the mma package \citep{s93}.
The bright quasar 3C273
was observed for flux calibration; its flux was derived from 
observations of either Uranus or Neptune taken from other observing sessions
in the array database. The flux of 3C273 changed slowly during the 12
months over which these data were obtained. The following values were
derived for the flux of 3C273: 21 Jy (1998 March-May);
22 Jy (1998 September - November 1); 19.5 Jy (1998 November 1 - December 20);
17 Jy (1998 December 20 - 1999 February 16). The absolute
calibration uncertainty is estimated to be 20\%.
The gain and passband calibrator for these observations was the bright 
quasar 3C279. Only observations for which the coherence on 3C279 was
measured to be $>50\%$ were included in the final analysis.

The data were mapped using the Miriad analysis package \citep{stm95}. The
{\it uv} data were first clipped to have a maximum amplitude of 14 Jy. The
three fields were inverted together with robust weighting to make a single
mosaic dirty map. The rms noise in this dirty map in line-free channels
was 0.055 Jy beam$^{-1}$ and the synthesized beam was
$3.15 \times 4.91^{\prime\prime}$ (or $310 \times 480$ pc at a
distance of 19 Mpc). The map was cleaned using the task MOSSDI with
a clean cutoff of 0.11 Jy beam$^{-1}$ (2$\sigma$). We used a
single clean box made by summing three
$60\times60^{\prime\prime}$ clean boxes centered on each of the three field
centers. As a final step, individual maps and the data cube were 
corrected for the fall-off
in sensitivity due to the primary beam of the 10.4 m antennas. We
used the task MOSSEN to create a map of the gain due to the primary
beam and then divided individual maps by that gain file, while also
masking regions of the maps where the gain map was smaller than 0.5.

\subsection{Identifying Clouds from the Data Cube}

Individual clouds were 
identified from the data cube using the automatic clump identification
algorithm CLFIND \citep*{w94}.
This algorithm searches for peaks of emission within a contour map
of the data, which it then follows down to lower intensity levels,
and has the advantage of not assuming any specific clump profile
(such as a gaussian profile). The main free parameter in the algorithm
is the contour level, which defines our cloud detection threshold. 
The contour level is specified relative to the noise
in the map, and is usually set to 2$\sigma$. Only emission
brighter than one contour level can be included in a cloud, and 
a cloud must have at least one pixel that is twice the minimum
contour level in order to be found.
Thus, the initial cloud identification was
run on the data cube before correction for primary beam attenuation.
We experimented with three different contour levels to see how
the contour level affected the properties of the clouds; we used
contour levels of 0.10, 0.11 ($2\sigma$), and 0.12 Jy beam$^{-1}$.
Unlike \citet{w00}, individual clouds were not inspected and
merged by hand where they appeared to overlap; thus, the properties
of the largest clouds will be somewhat different from the values in
\citet{w00}.

The properties of the individual clouds such as
position, flux, and velocity were measured using the program CLSTATS
\citep{w94}. This program combines the clump
assignment cube produced by CLFIND with the original data cube to
calculate the parameters of the individual clouds from the pixels
that have been assigned to that cloud. We used the gain-corrected
data cube in running CLSTATS to produce fluxes that were corrected
for the primary beam. The properties of the clouds measured
with a contour level of 0.11 Jy beam$^{-1}$ are given in 
Appendix~\ref{appendixA}.

To estimate the sensitivity of the flux of an individual cloud (and
hence its mass) to the chosen contour level, we took a closer look
at the largest clouds in our sample, those with masses estimated
to be larger than $10^8$ M$_\odot$. Nine large clouds
identified with a contour level of 0.11 Jy beam$^{-1}$ were cross-identified
with the lists from the other two contour levels, and the masses
measured for each of the three identification runs were compared. 
The average dispersion in 
the mass measurements determined from the three
different identification runs was 25\% while the dispersion for an 
individual clouds ranged from a low of 0 to a high of 55\%. Thus,
we estimate the random uncertainty on the mass estimate for any individual
cloud in our sample to be 25\%.

\subsection{Moment Maps and Total Flux}

The integrated intensity map presented in \citet{w00} 
was made using the command CLPLOT to sum 
together the emission in the data cube that corresponded
to molecular complexes identified by CLFIND.
Only molecular
complexes with emission in at least three velocity channels were
included in the integrated intensity map. 
Thus, this image (shown again in Figure~\ref{fig1}a) 
differs from a standard zeroth moment map 
because very small clouds and isolated, weak emission regions
have not been included\footnote{The 
version of CLPLOT that was used in this analysis had a bug in
that the last cloud of a list of 20 was not included in the plots.
This bug resulted in one small cloud not being included near
the southern end of the west ring and some small changes
to the contours around the two nuclei. The corrected version of the
plot is shown in Figure ~\ref{fig1}b.}. An integrated intensity map 
(not shown) produced in the
same way but including {\it all} the complexes identified by CLFIND, no
matter how weak, is very similar to the positive contours in
the zeroth moment map made
with a $2\sigma$ intensity cutoff (Figure \ref{fig1}c).
Figure \ref{mom1} shows the first moment map of the molecular complexes
in the Antennae. The original data cube was filtered by keeping only
emission that CLFIND identified with a molecular complex; the first
moment map was produced from this filtered data cube with no additional
flux cutoff. Thus, this first moment map includes emission down
to a level of  $2\sigma$, but is somewhat cleaner than a first moment
map made with a $2\sigma$ cutoff from the original data cube because isolated
emission regions extending over fewer than four pixels have not been included.

The zeroth moment map shows clear negative bowls 
(which are not shown in Figure \ref{fig1} for clarity) around the bright
emission regions, which are a sign that we are not detecting all
the flux in the region with the interferometer. These negative bowls
make it difficult to measure the total flux, so the total flux was
measured instead from the integrated intensity image from CLPLOT
which included the very small clouds. The integrated intensity
measured in this way is 930 Jy km s$^{-1}$, compared to 910 Jy
km s$^{-1}$ measured in \citet{w00}.
This flux is probably a slight underestimate of the total flux in the map,
since CLFIND sometimes misses weak emission at the edges of clouds.
To correct for this effect, we calculated the difference between
the CLFIND map and the zeroth moment map, and measured an additional
flux of 125 Jy km s$^{-1}$ 
in the two nuclei and the overlap region where significant
positive signal was seen. Thus, the final estimate of the total integrated
intensity detected with the interferometer is 1055 Jy km s$^{-1}$.
Adopting a 
CO-to-H$_2$ conversion factor of
$3 \times 10^{20}$ H$_2$ cm$^{-2}$ (K km s$^{-1}$)$^{-1}$ 
\citep{s88} and 
including a factor of 1.36 to account for heavy elements, this intensity
corresponds to  $6.1 \times 10^9$ M$_\odot$ of molecular gas. 
In Appendix~\ref{appendixB}, we estimate 
the CO-to-H$_2$ conversion factor for the Antennae from a few
large clouds and find that it agrees with the Galactic value within 
the uncertainties.

\citet{g01} have made a complete single-dish map of the Antennae system.
They measure a total flux of 3172 Jy km s$^{-1}$, which, for
the same distance and conversion factor adopted above, corresponds
to a total mass of $1.8\times 10^{10}$ M$_\odot$. This flux is
significantly larger than the flux detected in our interferometric map.
However, their map, which contains 73 spectra at 28$^{\prime\prime}$ spacing,
covers a substantially larger area than does our map. To estimate the 
single dish flux within our smaller map, we summed the spectral lines
from fourteen spectra which fell within the clean box used in processing
the interferometer map. The total flux from these fourteen spectra
is 1654 Jy km s$^{-1}$ for a mass of
$9.6\times 10^9$ M$_\odot$. Thus, the interferometer map appears to
have detected 65\% of the total flux seen over the same area in
the fully-sampled single-dish map. 

\section{Molecular Complexes in the Antennae and Other Galaxies \label{mass}}

\subsection{The Mass Spectrum of the Molecular Complexes}

We measured the mass spectrum for the molecular complexes in the
Antennae using the three different contouring levels discussed in
\S\ref{obs} to see whether the slope and upper-mass cutoff to the
mass spectrum were heavily dependent on the contour used to identify
the clouds. The total number of complexes identified was
114 for a contour level of 0.10 Jy beam$^{-1}$,
86 for a contour level of 0.11 Jy beam$^{-1}$, and
49 for a contour level of 0.12 Jy beam$^{-1}$.
For each set of complexes, we calculated the number of the clouds 
in logarithmic mass bins with a spacing of 0.2; two different centers
for the mass bins were also used, with the first having a bin centered at
$\log(M) = 7.0$ and the second having a bin centered at
$\log(M) = 7.1$. The number of clouds in each mass bin was then divided
by the central mass of that bin to obtain a true differential mass function,
$dN/dM$. The differential mass functions for the three different
identification runs are shown in Figure \ref{fig-diff-mass}.

We calculated the detection and completeness limits for our data using
a modified version of the formulae given in \citet{heyer01}. For these
interferometric data processed using CLFIND, a cloud must contain
at least four pixels, one of which exceeds twice the specified contour level,
in order to be detected. The detection limit in Jy km s$^{-1}$ is
then given by
$$S_{CO}^{min} = 5 \Delta S \Delta V / 17.55$$
where $\Delta S$ is the contour level in Jy beam$^{-1}$, $\Delta V$ is
the velocity width of a single channel (5.2 km s$^{-1}$) and the
factor of 17.55 is the area of the synthesized beam in pixels. For
$\Delta S = 0.11$ Jy beam$^{-1}$, the detection limit is
$S_{CO}^{min} = 0.16$ Jy km s$^{-1}$. The $5\sigma$ completeness
limit is given by 
$$S_{CO}^c = S_{CO}^{min} + 5\sigma(S_{CO})$$ 
where
$$\sigma(S_{CO}) = \sigma \Delta V \sqrt{N_pN_c/17.55}$$
and $\sigma$ is the rms noise in the map
$N_p $ is the minimum number of pixels required for a detection
and $N_c $ is the minimum number of velocity channels required for
a detection. For $\sigma =0.055$ Jy beam$^{-1}$, $N_p = 4$, and $N_c$ = 1,
we get $\sigma(S_{CO}) = 0.14$ Jy km s$^{-1}$, which gives a 
$5\sigma$ completeness limit of 0.86 Jy km s$^{-1}$ or
a mass limit of $5.0\times 10^6$ M$_\odot$.

We made least squares fits to the six mass functions including $\sqrt N$
uncertainties and ignoring bins with $\log(M)\le 6.7$. The slope
and uncertainty in the slope are included in Figure \ref{fig-diff-mass}.
The average slope across the three different identification
runs is $\alpha = -1.4\pm0.1$. This slope refers to a mass range
from $5\times 10^6$ M$_\odot$ to $\sim 10^9$ M$_\odot$ and 
is significantly shallower
than the canonical stellar initial
mass function derived by \citet{s55} (-2.35 on the same scale).
However, this 
slope of -1.4 
is very similar to the slope of -1.5 obtained for 
the mass function of Giant Molecular
Clouds in the Milky Way \citep{sss85,s87}. It is somewhat shallower
than the slope of -1.8 found over a mass range of 1000 M$_\odot$ to
$10^6$ M$_\odot$ for molecular
regions in the outer Galaxy \citep{heyer01}. However, \citet{heyer01} point
out that many of the low-luminosity objects in their survey are
not self-gravitating, in which case the mass function may be 
shallower than the luminosity function. In addition, 
the relatively poor spatial resolution of our Antennae data means that
blending of individual clouds into
a single feature could be an important effect, which would tend to
make our observed mass function shallower than the true mass function.
\citet{k98} also obtained 
similar slopes ranging from -1.6 to -1.8 from CO observations
of clumps inside seven molecular clouds. The clump masses in
their sample ranged from as low as $10^{-4}$ M$_\odot$ to as high
as $10^4$ M$_\odot$ and overlap at the high-mass end with the
mass range covered by \citet{heyer01}. 
[However, steeper clump mass functions, consistent
with the stellar initial mass function  over a fairly 
narrow mass range from $\sim 0.5$ M$_\odot$ to $\sim 10$ M$_\odot$, 
have been derived recently
for several nearby clouds using submillimeter continuum data \citep{ts98,mo98,
j00,m01}.] 
Thus, by combining our new
study of the Antennae with previous work on the Milky Way, we 
see that the mass function of structures in the molecular 
interstellar medium in galaxies 
has a fairly constant slope of $\sim -1.5\pm0.2$
that extends over at least 8-9 orders of magnitude in mass,
from 1-10 $M_\odot$ up to $10^9$ M$_\odot$. It is worth noting
that these power-law mass spectra are consistent with a model
where clouds grow by agglomeration from smaller objects
\citep{k79}.

\subsection{Previous Observations of 
Molecular Complexes in Other Galaxies}

We have searched the literature for CO observations of galaxies with
spatial resolution of better than 1 kpc to see what is known about
massive molecular complexes in other galaxies (Table \ref{tbl-other-gmas}).
Objects similar to the most massive complexes in the Antennae have
been found in only three other galaxies: NGC 1068
\citep{p91}, Arp 220 \citep{sa96}, and Arp 299 \citep{c99}. 
Adopting a distance of 14.4 Mpc for NGC 1068 \citep{tf88},
the most massive complex is $4.5\times 10^8$
M$_\odot$ \citep{p91} 
and the average molecular gas surface density in the inner
arcminute is roughly 100 M$_\odot$ pc$^{-2}$, both comparable to what
is seen in the Antennae.
In Arp 220 and Arp 299,
the massive gas concentrations are located in the galactic
nuclei, which probably makes them more similar to the gas located in
the nucleus of NGC 4038 than the molecular complexes in the disks and
overlap region. In the remaining normal spiral galaxies for
which high resolution CO observations have been made, the most
massive complex ranges from as small as $1.6\times 10^7$ M$_\odot$
in M83 \citep{r99} to as large as $1.3\times 10^8$ M$_\odot$ in
M100 \citep{r95}. 
For the five galaxies in Table \ref{tbl-other-gmas}
with distances between 9 and 19 Mpc, the angular resolution of the
data sets varies by only a factor of two, while the mass of the most
massive complex varies by a factor of almost twenty. This result
suggests that the large complexes seen in the Antennae and in NGC 1068
are not the result of cloud blending due to the relatively
large distances of these two galaxies.

In only two galaxies, M51 \citep{rk90} and NGC 1068 \citep{p91}, are
sufficiently large numbers of objects detected that it is possible
to look at the mass function of the molecular complexes. However,
the limited mass range of the molecular complexes in M51 ($1-6\times 10^7$
M$_\odot$) makes it impossible to determine the slope with any degree
of accuracy. The 38 complexes detected in NGC 1068 range from $2\times 10^7$
M$_\odot$ to $7\times 10^8$ M$_\odot$. Fitting the entire mass range
in the same manner as was done for the Antennae data gives a slope of
$-1.3\pm0.2$ for the differential mass function, which is
quite similar to
that obtained in the Antennae.
From this comparison with previous studies of other galaxies, it is clear
that the population of molecular complexes identified in the Antennae is
both the largest sample and covers the largest dynamic range in mass.
It would be interesting to carry out deeper observations of
Arp 220 and Arp 299  to see if the mass functions of the molecular
complexes in these more advanced interacting systems 
are similar to those measured for the Antennae and NGC 1068.

\subsection{The Spatial Relationship Between Young Massive Star
Clusters and Super Giant Molecular Complexes in the Antennae\label{SC-GMC}}

\citet{w99} have published a large sample of young massive star clusters
observed with WFPC2 on the {\it Hubble Space Telescope}. 
They find that the overlap region contains
primarily clusters with ages $< 5$ Myr. This region contains roughly half
the total amount of molecular gas detected in our map and has correspondingly
very high extinctions, with an average $A_v$ of 96 mag towards the three
brightest CO peaks in this region. Thus, any clusters identified in
this part of 
the overlap region must lie on the near side of the molecular complexes.
The western loop of gas and young stars to the west of NGC 4038 contains 
clusters with ages primarily in the range of 5-10 Myr. In this region,
the CO emission occurs primarily between regions of many young clusters
(Figure \ref{optcoplot}),
and the typical extinction towards a cloud in this region is
$A_v \sim 13 $.
A very interesting region from the point of view of cluster formation
is the north-eastern portion of NGC 4038. \citet{w99} find
that this region contains many young massive star clusters, with two-thirds of
the clusters having ages less than 30 Myr and one-third having ages
around 100 Myr. The presence of the younger star cluster population is
interesting given the total lack of detected CO emission in this part
of our map.
Our 5$\sigma$ detection limit of $5\times 10^6$ M$_\odot$ (\S\ref{mass})
means that only relatively normal giant molecular clouds can be present
in this region of the galaxy.  

\citet*{z01} have compared the locations of three age groups of star
clusters with data at a variety of other wavelengths. They find that
clusters with ages $< 5$ Myr (the reddest clusters) are more associated
with wavelengths longer than mid-infrared, while clusters with ages
$> 10$ Myr are more closely associated with far-ultraviolet and X-ray
emission. The 2-point correlation function for the red clusters is
a power-law up to a radius of 0.74 kpc. The maximum size of the power-law
portion of the correlation function is very similar to the sizes of
the super giant molecular complexes, 
which suggests that an upper limit to the correlation function
could correspond to a set of clusters formed inside a single such complex.
\citet{z01} estimate a total star formation rate from H$\alpha$ emission
(corrected for 1.8 magnitudes of extinction and assuming a 
Salpeter initial mass function from 0.1 to 100 M$_\odot$) 
of 20 M$_\odot$ yr$^{-1}$. 
This star formation rate is a factor of two larger than the star
formation rate of 10 M$_\odot$ yr$^{-1}$ which can be 
derived from the far-infrared luminosity following \citet{ken98}.

We compare the locations of the young, red star clusters discussed
in \citet{z01} with the distribution of the molecular
gas in Figure \ref{optcoplot}. Roughly 40\% of these young, obscured clusters
lie within the region of strong CO emission in the overlap region, which
occupies only 10\% of the total area mapped in CO. Thus, we 
confirm the conclusions of \citet{z01} 
that the youngest clusters are significantly correlated with 
regions containing molecular gas. However, some of these youngest
star clusters lie a significant distance from strong CO emission, up to
1-2 kpc in regions to the north and east of the galaxies. 
This result is also true for the older B1 and B2 cluster groups
[compare Figure \ref{optcoplot} with Figure 2 of \citep{z01}]. 
Since our
CO map does not contain all the CO flux of these galaxies, it is possible
that significant molecular gas exists near these otherwise isolated
young clusters. However, this molecular gas must be in rather low-mass clouds
to have avoided detection in our map; in particular, any remaining
clouds associated with these clusters are likely to have masses less
than $5\times 10^6$ M$_\odot$, our 5$\sigma$ detection limit for clouds.
Comparing the CO and cluster distributions thus suggests that
either relatively low-mass molecular clouds can form the occasional
massive star cluster, or that massive star clusters can destroy their parent
clouds on a rather rapid time scale, or both.

\section{A Case Study Comparison with M51 \label{M51comp}}

To illustrate better the unusual aspects of the molecular interstellar
medium in the Antennae, we compare the total gas mass and the properties
of the molecular complexes and young
massive clusters to the grand-design spiral galaxy M51.
(Unfortunately there seem to be no surveys for young massive star
clusters in NGC 1068, which would otherwise be another interesting
galaxy to compare to the Antennae.)
\citet{rk90} mapped a $4\times 5^\prime$ region of the inner disk of M51 with
the three-element Caltech Millimeter Array and detected
26 giant molecular associations with masses ranging from
$1\times 10^7$ M$_\odot$ to $5\times 10^7$ M$_\odot$. The beam size ($9\times 
7^{\prime\prime}$ or $370\times 290$ pc at a distance of 8.6 Mpc, 
\citep*{f97,f00})
is a very good match to our study of the Antennae. 
To match the total area surveyed
as well, we restrict our analysis to clouds within a
5 kpc radius of the nucleus of M51. 
Correcting for the larger distance to the Antennae,
our study is 1-2 times more sensitive to a complex of a given mass
than is the M51 study of \citet{rk90}. To correct for this difference
in mass sensitivity, we further restrict our comparison to complexes
that have masses larger than $2\times 10^7$ M$_\odot$.
More recent work by \citet{a99} provides better resolution and
sensitivity for M51, but we prefer to use the older study of 
\citet{rk90} because it provides a better match in spatial resolution
and mass sensitivity to our Antennae data.

M51 contains 13 molecular complexes with masses greater than $2\times 10^7$ 
M$_\odot$ inside an area of 79 kpc$^2$.
In contrast, the Antennae contain 43 complexes with masses
greater than $2\times 10^7$ M$_\odot$ inside the same area. Thus, the
Antennae have a three times higher surface density of massive
molecular complexes than does M51. (However, correction for the
poorly known inclination of the disks of NGC 4038 and NGC 4039 might
reduce this ratio by up to a factor of two.)
In addition, the most massive complex
seen in M51 has a mass of $5\times 10^7$ M$_\odot$, while the most
massive complex in the Antennae (ignoring the gas in
the nucleus of NGC 4038) has a mass of $6\times 10^8$ M$_\odot$,
12 times more massive than what is seen in M51. Thus, while the
Antennae have a somewhat higher surface density of massive molecular complexes
than does M51, the most striking aspect of the molecular complexes
in the Antennae is that their mass function extends to much
larger masses than are seen in M51. 
If the molecular complexes in both galaxies follow a similar power-law
mass function, then fitting the data with truncated power-laws
following the prescription of \citet{s01} shows that the
M51 mass function is significantly truncated. In particular,
given that 13 complexes are observed with masses greater
than $2\times 10^7$ M$_\odot$, we would expect to see
4 complexes with masses greater than $6\times 10^7$ M$_\odot$.
Thus, it is probably not small number statistics
that is imposing the upper mass
limit in M51.

It is important to compare 
the total amount of molecular gas in the Antennae and M51 to see
whether the presence of very massive complexes can be attributed
to a larger total reservoir of molecular gas to form them.
Within the region of our study, \citet{g01} measure
a total mass of $9.6\times 10^9$ M$_\odot$ (\S\ref{obs}). This mass
corresponds to a surface density of 120 M$_\odot$ pc$^{-2}$ (again
with no correction for inclination, which could reduce this
value by up to a factor of two). 
\citet{g93}
made a large map of M51 using the IRAM 30 m telescope; within
the central 79 kpc$^2$ and using a conversion factor that
is three times smaller than the value adopted here for the Antennae,
their measurements correspond to a total mass of molecular
gas of $6\times 10^{9}$ M$_\odot$ or a surface density of
90 M$_\odot$ pc$^{-2}$. Thus, within 
uncertainties due to inclination in the Antennae and the CO-to-H$_2$
conversion in M51, the total molecular gas surface density appears
comparable in the two galaxies and certainly not different enough
to account for the formation of extremely massive molecular complexes
in the Antennae.

One important difference between the two galaxies is their velocity fields.
Figure~\ref{mom1} shows that the Antennae have
a reasonably regular velocity field
in the north-western arc, which may represent a fairly unperturbed
piece of the disk of NGC 4038. However, the velocity field
in the overlap region appears extremely disturbed; instead of a smooth
progression in velocity, there are two large regions with fairly
constant velocities separated by a boundary with a steep velocity gradient.
In contrast to the disturbed velocity field of the Antennae system,
the velocity field in M51 shows evidence of 
streaming motions due to the presence of spiral density waves
superimposed on a very regular rotation pattern \citep{a99}. This
marked difference in the velocity fields in M51 and the Antennae
(which is not unexpected given the strong interaction in the Antennae)
suggests a possible explanation for the presence of very massive
molecular complexes in the Antennae. 
%**vv
In a regularly rotating disk,
molecular clouds that contain enough mass to be gravitationally bound
may be subject to disruption by the effects of tides and shear
\citep{r93}. Thus, it is possible that
the formation of very massive gas clouds is suppressed in M51
due to high shear in its differentially rotating disk. The effect of
galaxy mergers like the Antennae 
on the shear in the galactic disks is not well
understood, althought it seems possible that some regions undergo
increased shear while other regions may experienced reduced shear.
It is interesting that for NGC 1068 
(the other nearby spiral galaxy with massive
gas clouds), \citet{s00} have suggested that
the gaseous spiral arms may lie at the Inner Lindblad Resonance of
a much larger (17 kpc) bar. Thus, the massive gas clouds in NGC 1068 
may also have formed in a region of reduced shear.
%**^^

Surveys for young massive star clusters in M51 have been carried out recently
by \citet{l00} and \citet{l02}.
Using ground-based imaging,
with HST archival confirmation of 10 clusters, \citet{l00} identified a 
population of 69 candidate young clusters with ages less than 500 Myr.
This sample contains 40 clusters with $M_V < -10$~mag and 4 clusters with 
$M_V < -12$~mag.
\citet{l02}
used HST WFPC2 imaging of the nuclear region of M51
to identify a sample of 30 point-like sources. However, they concluded that
these sources were most likely to be single stars or very small ($M < $ few 
100 M$_\odot$) star clusters, rather than massive star clusters. 
At the extremely young end of the scale, \citet{s01}
concluded from their Pa$\alpha$ and H$\alpha$ surveys that the 
most massive star cluster in M51 that is ionizing an HII region has
a mass $< 5000$ M$_\odot$. \citet{l00} notes that M51 is rich in young massive
star clusters compared to an average spiral galaxy, but that it is not
{\it unique} in its rich star cluster population.

In comparison, in just a single 10 kpc$^2$ region of the Antennae,
\citet{w99}  find a total of seven clusters with
$M_v < -12$~mag. Although this region of the Antennae is particularly
rich in clusters, it seems likely that the Antennae as a whole contain
an order of magnitude more of the most luminous clusters than does M51.
A total of 14,000 point sources have been identified in the Antennae
\citep{w99}, of which at least 40\% are contaminating stars and
at least 800 are definitely star clusters. \citet{zf99} identify
2000 star cluster candidates from this sample which lie within
0.3 mag of cluster evolutionary tracks and have $M_{v,o} < -9$ mag.
Compared to
the $\sim 70$ star clusters identified in M51 by \citet{l00}, these
observations
suggest that the Antennae 
contain at least 30 
times as many star clusters
as M51. This difference is even more striking when we consider the
average star formation rates in the two galaxies. The star formation
rate is roughly proportional to the far infrared luminosity 
(i.e. \citet{ken98});
M51 has a far-infrared luminosity of $1.8\times 10^{10}$ L$_\odot$
\citep{r88}, while the far-infrared luminosity in the
Antennae is $5.6\times 10^{10}$ L$_\odot$ \citep{g01}, roughly
3 times larger than M51. Thus, in comparison to their star formation
rates, the Antennae contain an order of magnitude more young massive star
clusters than does M51. We suggest in \S\ref{SSCF} that the
formation of young massive clusters in the Antennae is enhanced by
its ability to form very massive gas clouds.

\section{Is the Antennae Undergoing a Short-Lived Starburst?
\label{short-lived}}

\subsection{Recent Star Cluster Formation Rates from Optical and Radio Data}

We can derive clues to the recent star cluster formation rate in the
Antennae by examining the number of massive star clusters as a function of
age. We can also use radio continuum observations to estimate the number
of very young, embedded star clusters 
and older supernova remnants \citep{nu00}. 
The mass function analysis of \citet{zf99} shows
that there are roughly 300 young star clusters more massive than 
10$^5$ M$_\odot$
in the Antennae. Of these clusters, roughly 100 fall in the young age
range of 2.5-6.3 Myr and a similar number fall in the age range of 25-160 Myr.
There are two simple explanations for the similar number of
star clusters observed in these very different size age bins. 
If the observed number of star clusters
reflects the average cluster formation rate over each time period, then 
the cluster formation rate in the last 6 Myr must have 
been 20 times higher than
the cluster formation rate over the previous 150 Myr. However, this picture
implies a very rapid change in the cluster formation rate compared
to the current crossing time of the system ($\sim 100$ Myr, 
\citet*{m93}). Alternatively, if the
cluster formation rate has been constant over the last 150 Myr, then
only roughly one out of twenty of the younger clusters can survive in 
the long term \citep{zf99,w02}. 
\citet{m02} have measured dynamical masses for five bright clusters
in the Antennae, which suggest that these clusters are gravitationally
bound and hence likely to survive. However, we
have no information about the likelihood of survival for the vast majority
of the star clusters in the Antennae. A counter example of a luminous
star cluster that seems to have insufficient low-mass stars to survive
can be found in M82 \citep{sg01}.

A recent high-resolution radio survey shows that 
the Antennae contain
13 sources with obviously thermal spectral energy distributions
which have likely ages in the range of 1-3 Myr \citep{nu00}.
These sources contain an equivalent mass in stars between 0.1 and
60 M$_\odot$ (assuming a Salpeter initial mass function) of $2 \times 
10^5$ to $1\times 10^7$ M$_\odot$ 
\footnote
{These masses can be obtained by scaling the mass in O5 stars by a factor
of 45; assumes ionizing luminosity$L_i \propto m^{3.5}$,
$L_i({\rm O5}) = 4.7\times 10^{49}$ erg s$^{-1}$, and
$M({\rm O5})= 60$ M$_\odot$; Ulvestad, private communication.}.
Roughly accounting for 
sources that have been missed because of large uncertainties
on their spectral index as well as slightly less massive
sources, we estimate that the number of optically thin thermal sources
with masses greater than $10^5$ M$_\odot$ is perhaps 30. 
In comparison, there are roughly 100 star clusters more massive
than $10^5$ M$_\odot$ with ages $\le 6$ Myr \citep{zf99}.
Thus, the number of optically visible star clusters
and the number of optically thin thermal sources appear to be consistent with
a constant star cluster formation rate in the Antennae over the last
6 Myr. 

These numbers are also consistent with the result
from \citet{wz02}, who find that essentially all the bright thermal
radio sources can be identified with optically visible star clusters.
This good match between the radio and optical sources is
rather puzzling in the overlap region, where the visual extinction
can reach as high as 100 mag (\S\ref{SC-GMC}).
However, a closer examination of the sources in the overlap region reveals
that all sources except WS-80 lie well away from the brightest CO peaks. 
WS-80 itself is coincident with SGMC 7, which has an average visual
extinction of 25 mag. \citet{wz02} estimate a visual extinction for
WS-80 of 7.6 mag, which is consistent with it lying fairly close to
the center of this somewhat lower mass molecular cloud. 

Turning now to the supernova rate, which traces star formation over
slightly longer time periods, \citet{nu00} used the observed
flux in non-thermal radio sources to estimate a current
supernova rate of 0.1-0.2 yr$^{-1}$. This rate is a factor of 10-20 larger
than the rate of 0.01 yr$^{-1}$ they estimate from the number of O5 stars
and the lifetime of these stars. This result 
led them to postulate that the Antennae
suffered a very sharp burst of star formation 3-4 Myr ago that lasted
only a few hundred thousand years. However, the 60 M$_\odot$ O5 stars are
not the only stars that will produce supernovae; conservatively, all stars
with masses greater than 20 M$_\odot$ will produce a supernova. There will
be many more of these lower mass O stars but they will have significantly 
longer main sequence lifetimes. However, if the star formation rate has
been constant over the longer lifetimes of these stars, then the numbers
of O stars will approach a steady state, with the number of stars at
any mass proportional to the product of the initial mass function and
the lifetime of the stars. This effect will  increase the 
supernova rate over that calculated by \citet{nu00}. For
example, if the stellar lifetime were to vary as $t \sim m^{-1.65}$, then 
the number of supernovae produced by 20-50 M$_\odot$ stars would be 35 times
the number produced by 50-60 M$_\odot$ stars. Thus, depending on
the exact relationship between mass and lifetime and the lower mass
limit for producing a supernovae, it appears that the observed 
supernova rate in the Antennae is roughly consistent with a constant
star formation rate over the last 10-20 Myr.

Given the results from the comparison with the radio data, it seems reasonable
to assume that the star cluster
formation rate over the last 160 Myr was roughly constant, but that only a 
small fraction of the clusters formed 160 Myr ago 
survive to the present time \citep{fz01}.
This scenario would involve forming $\sim 2000$ 
clusters over the last 160 Myr with a total initial mass 
in the clusters of $\sim 10^9$ M$_\odot$. This large mass is still only a 
small fraction of both the total mass of molecular gas (Gao et al. 2001) and
the total luminous mass in the Antennae. 
Thus, the amount of gas available to fuel star formation in the Antennae
does not require that the current epoch of intense star cluster formation 
be confined to a very short period of time. A model
with a roughly constant cluster formation rate is consistent
with a picture in which the starburst activity in the Antennae
is dynamically triggered, in which case we would not expect the
star formation rate and star cluster formation rate to change
on a much shorter timescale than the dynamical time, or on
the order of 100 Myr \citep{m93}.

\subsection{The Future Rate of Cluster Formation in the Antennae}

We estimate the {\it future} rate of star
cluster formation in the Antennae using a model
based on our knowledge of current star formation in the Milky Way.
If super giant molecular complexes are similar in their star 
formation properties to giant molecular clouds
in the Milky Way, then we might assume that they will form at
most 5\% of their mass into stars. 
By analogy with what is seen in the Orion molecular cloud
\citep{l91a}, we would expect
these stars to form predominantly in star clusters.
Given the total mass in super giant molecular complexes is
$6\times 10^9$ M$_\odot$, we might expect them to
form a total mass of star clusters of up to
$3\times 10^8$ M$_\odot$. 
\citet{zf99} find that the mass function of the young massive
clusters in the Antennae follows a power law with a slope
of -2. This mass function implies that there is
an equal amount of mass in equal logarithmic
mass bins and thus the number of clusters more massive than
$10^5$ M$_\odot$ depends on the lower mass cutoff to the mass function.
For example, 
if star clusters form over a mass range from $10^4$ to $10^6$ M$_\odot$,
then the mass function will contain 600 clusters more massive
than $10^5$ M$_\odot$, whereas if star clusters 
form over a mass range from 100 to $10^6$ M$_\odot$,
the number of massive clusters
would be only half as large. These rough estimates suggest that
we might expect the existing super giant molecular complexes
in the Antennae to form between 300 and 600
 star clusters more massive than $10^5$ M$_\odot$.
These clusters will likely form over at least 
a crossing time, which for the largest complex is 30 Myr. 
This implies a massive star cluster formation rate of
10-20 Myr$^{-1}$,
which is approximately the same rate as is seen in the
youngest star cluster sample of \citet{zf99}. 

If  clusters older than 30 Myr have been subject to destruction, it
is possible that the Antennae have been experiencing a high and uniform rate of
star cluster formation over the last 100-200 Myr and could 
possibly sustain this formation rate for at least another 30 Myr.
What might happen after that? If the molecular gas can reform itself into 
super giant molecular complexes
(or other structures) capable of forming massive star clusters over that same
timescale, then massive star cluster formation could continue at the
same rate until all the gas is exhausted.
However, the north-east region of NGC 4038 suggests that, once a region
has formed many young massive 
star clusters, it is unlikely to be ready to form them
again just 30 Myr later. On the other hand, the fact that we see star clusters
forming now in the overlap region suggests a mode of star formation where a 
portion of the galaxy lights up in young clusters formed 30 Myr ago, while 
another portion
of the galaxy is forming the next generation of clusters. Perhaps the western
arc represents an intermediate age region where we still see gas and 
star clusters
intermingled. In this picture, a continuous period of star cluster formation
extending well in excess of 
30 Myr but isolated to only a portion of the galaxy at 
a time could perhaps be sustained. If this mode of cluster formation
continued for as long as 600 Myr, the Antennae would form 6000-12,000
clusters, of which as few as 300-600 might survive in the long term.
Slow infall of atomic material from the tidal tails \citep{h01,hm95}
could help prolong the period of active star cluster formation.

We need to consider whether
the expected total star cluster formation rate in this
simple model would cause the Antennae to exceed the typical
specific frequency for globular (and other) clusters in an elliptical galaxy. 
The current $V$-band absolute magnitude of the Antennae system is about -22 
(from data in LEDA). 
The ``normal'' value for the specific frequency is about
3.5 \citep{harris01}, which means we would expect to find
a total of about 2200 clusters more massive than $10^5$ M$_\odot$
in an elliptical galaxy with $M_v = -22$~mag. If the
Antennae were to fade by 1.5 mag in the process of forming an
elliptical \citep{ws95}, then the expected number of clusters would be 600.
Whether these numbers agree in detail  with the model discussed above
depends upon both the duration
of cluster formation and how many of
the young massive star clusters survive over the long term. There is
no difficulty if 
only 5-10\% of the young massive star clusters that could be formed
in future from the molecular gas reservoir survive as clusters
after a Gyr. However, if all the young (2-6 Myr) massive
star clusters that we see now will 
survive to old age, we would have to be witnessing an
extremely short-duration starburst in order for the final
specific frequency of the Antennae not to exceed (by a wide margin!)
the typical specific frequency seen in elliptical galaxies.
In this context, it is relevant that somewhat older
($\sim 1$ Gyr) mergers such as NGC 3921 \citep{s96} and NGC 7252
\citep{m97} are likely to end up with specific frequencies that fall in the
normal range of field elliptical galaxies.

\section{Forming Massive Star Clusters: Constraints from the Gas Properties
\label{SSCF}}

\subsection{Two Contrasting Models for Massive Star Cluster Formation}

\citet{lr00} show that the fraction of the U-band luminosity
that originates in young star clusters is proportional to the star formation
rate per unit area in a galaxy. Roughly speaking, their result suggests
that if you double the surface density of star formation, you double the
fraction of the galaxy luminosity that occurs in star clusters. 
The luminosity in star clusters depends in a complicated way on the star
formation history of a galaxy as well as on the survival rate of the star
clusters. However, using the U-band luminosity function will preferentially 
select clusters that are typically $< 30 $ Myr old, which is comparable
to the timescale over which the star formation rate is traced via
H$\alpha$ imaging. \citet{l00}
concludes that galaxies form young massive star clusters 
wherever the star formation rate is high enough. He suggests that
this could be linked to the formation of Giant Molecular Associations,
which may assemble more easily in galaxies with a higher gas surface
density. However, our comparison of M51
and the Antennae in \S\ref{M51comp} shows that they have similar 
gas surface densities and yet have very different star cluster
populations, 
and so gas surface density cannot be the only important parameter in the 
formation of young massive star clusters.

There have been various models proposed to explain the formation of young
massive star clusters. \citet{s96}
has noted the similarity between the
masses of the young star clusters and the masses of giant molecular clouds
in the Milky Way ($10^4 - 10^6$ M$_\odot$). However, the radii of
the two objects are quite different, with the young massive
star clusters having
effective radii of 4 pc (Whitmore et al. 1999) and giant molecular
clouds having radii in the range of 5 to 50 pc, depending on their
mass. \citet{s96} have suggested that a young massive 
star cluster could be formed through shock compression of a giant molecular
cloud to force the cloud to form stars with an efficiency of $\sim 50\%$, 
much higher
than the star formation efficiencies of a few percent that are typically
seen \citep{el91}. They suggest that one 
means by which giant molecular clouds
could possibly be compressed is through the formation of a hot, high-pressure
interstellar medium
during galaxy collisions \citep{js92}. This model was 
originally developed to explain the formation of a starburst during galaxy
interactions. In the model, the outer layers of a giant molecular cloud
are compressed by contact with hot, shocked HI clouds and form stars with
a high efficiency. However, the amount of the cloud mass that is subject to
this high efficiency is estimated to be 10\%, and would be distributed in
a shell-like structure rather than a compact, spherical structure. In addition,
\citet{js92} assume that only stars more massive than 1 M$_\odot$ are
formed in this process in order to match the observed parameters of
starburst galaxies, in particular their high
infrared luminosity. Thus, this model does not seem to be directly
applicable to the formation of compact massive star clusters.

An alternative model that has been proposed to explain the formation
of globular clusters \citep{hp94,mp96}
may be applicable to the formation of young massive 
star clusters in galaxy mergers.
This model supposes that globular clusters formed by the same type of
process that we see forming star clusters today. However, the large masses
of globular clusters compared to star clusters in clouds like Orion
requires that the globular clusters form within molecular clouds that are
substantially larger than the molecular clouds in the Milky Way. For
example, the Orion B molecular cloud has a total mass of $8\times 10^4$
M$_\odot$ and contains five molecular cores with masses ranging from
100 to 450 M$_\odot$ \citep*{l91b}.
Four of these cores are
forming infrared star clusters with a star formation efficiency of
50\% \citep{l91a} and most of the star formation is occurring 
in these cores. If a globular cluster
(or a young massive star cluster) formed inside a larger molecular core with
a star formation efficiency of 50\%, this would require cores
with masses from $2\times 10^5$ to $2 \times 10^6$ M$_\odot$. If these
large cores were contained within a larger cloud with a core-to-cloud 
mass ratio 
similar to Orion B, the cloud mass required to contain these cores would
be $4\times 10^7$ to $4\times 10^8$ M$_\odot$. These cloud masses are
very similar to those of the super giant molecular
complexes we have identified in the Antennae. Note
that in this model we would expect each super giant molecular
complex to contain several massive
cores and hence to form several massive star clusters over its lifetime.

\subsection{The Role of Pressure in the Two Models}

One possible way to distinguish between these two models may be
via the environment in which the clouds live, particularly the pressure.
We can estimate the pressure in a few regions in the Antennae from the
recent work by \citet{f03}.
They have made Chandra observations
which are sufficiently sensitive to be able to fit the physical conditions
in the hot interstellar medium in the two nuclei and also in a region in the western
star-forming arc. Assuming a scale height of 200 pc for the high temperature
component, the pressure in each of these regions is in the range 
of $4-8 \times
10^5$ K cm$^{-3}$, with an average over the three regions of $6\times 10^5$
K cm$^{-3}$. This pressure is significantly larger than the typical
interstellar medium pressure in the Milky Way of $10^4$ K cm$^{-3}$
\citep{e89}. 
If the pressure in
these three regions is typical of the Antennae disk as a whole, then the
super giant molecular complexes 
in the Antennae are embedded in a higher pressure environment than are
molecular clouds in the Milky Way. 

How does this observed pressure compare to the models of star cluster formation
discussed previously? \citet{hp94} adopt a surface pressure
for molecular clouds that is ten times larger than the typical interstellar medium pressure
to account for the effect of embedding HI envelopes. 
Using the formula in \citet{e89} and
HI data from \citet{h01}, the pressure due to the average atomic gas column
density in the Antennae is only $\sim 10^4$ K cm$^{-3}$, closer to
the Milky Way pressure than any other component of the interstellar medium in
the Antennae. Thus, 
the surface pressure felt by the super giant molecular complexes
in the Antennae is likely to be in the range of $6\times 10^5$ to $6 \times 
10^6$ K cm$^{-3}$, depending on whether they are surrounded by massive
HI envelopes with pressures significantly above the average pressure
in the atomic gas. Since the radius of a cloud varies as $P^{-1/4}$, 
a cloud of $3\times 10^8$ M$_\odot$ should have a radius between 300 and 
600 pc. This size is a reasonable match to those of the molecular
complexes in the Antennae.
The model of \citet{js92} predicts a pressure of $10^8$ K cm$^{-3}$
in the hot, shocked HI clouds. Although this pressure is substantially higher
than that estimated from the X-ray data, the expected
filling factor of this hot gas is less than 1\%. With
this low filling factor, the average pressure
predicted in the model is $6\times 10^5$ K cm$^{-3}$, which is quite similar
to the X-ray data. Thus, it seems that the pressure of the interstellar 
medium in the Antennae cannot help us to
distinguish between the two  models.

Even the high pressures inferred from the X-ray data for the Antennae
are still substantially smaller than the typical pressure inside
a single, self-gravitating molecular complex. The average pressure in SGMC-3 is
$2\times 10^6$ K cm$^{-3}$, while the central pressure is
in the range of $10^6-10^7$ K cm$^{-3}$, depending on the exact geometry.
Indeed, the pressure
inside the $\rho$ Ophiuchus core,  a relatively small star-forming
cloud in the Milky Way, is $10^7-10^8$ K cm$^{-3}$ \citep{j00}.
\citet{ee97} estimate the pressure required to form a typical globular
cluster to be in the range of $10^6-10^8$ K cm$^{-3}$.

\subsection{Applying the Models to the Non-Merger Environment}

It is important to see if we can understand the formation of young
massive star clusters
in non-interacting galaxies in the context of these two
models. A particularly useful case study is the R136 star cluster in
the LMC. The total mass of stars in R136 is estimated to be
$6\times 10^4$ M$_\odot$ \citep{h95}
and the age of the
star cluster is at most 1-2 Myr \citep{mh98}.
There appears
to be no molecular gas in the immediate environment of R136 \citep{j98};
however, there is a very extended region of CO emission
to the south of R136 \citep{c88}.
If we consider this
entire region as a single large structure, it has a size of $ 400 \times 1200$
pc
and a total mass of about $9\times 10^6$ M$_\odot$ (adopting the
appropriate CO-to-H$_2$ conversion factor from \citet{w95}).
Adopting the velocity width of 10.5 km s$^{-1}$ measured by
\citet{k97},
the virial mass of this region is 
$1.5\times 10^7$ M$_\odot$
or within a factor of two of the flux-based mass. Thus, this large CO complex
may be gravitationally bound. If R136 formed in a core of $1\times 10^5$ 
M$_\odot$, 
we would expect this core to have lived inside a larger cloud of mass
$2\times 10^7$ M$_\odot$, which is similar to the mass of the
large CO complex. Thus, it seems reasonable to suggest that R136 might have
formed from the most massive core of that molecular complex. R136 and
30 Doradus are rather isolated at present from the main part of this
molecular complex. However, the intervening region contains a
number of HII regions and supernova remnants \citep{c88}, which could have
acted along with R136 and 30 Doradus to destroy much of the molecular
gas over this larger region. Thus, the environment of
the R136 cluster appears roughly consistent with the cluster formation
picture put forward by \citet{hp94}, while the over-pressure model advanced
by \citet{s96} seems less applicable here.

Returning now to M51, the CO data show that it currently contains 13 molecular
complexes with masses in the range of $2-5\times 10^7$ M$_\odot$
\citep{rk90}. Scaling again
from Orion, we would expect the most massive core in each complex to be in the 
range of
$10^5-3\times 10^5$ M$_\odot$ and hence to be able to form a 
young star cluster
with a mass up to $1-2\times 10^5$ M$_\odot$. The 69 star clusters identified
by \citet{l00} have ages $<$500 Myr. Depending on the masses of these star
clusters (which have not been estimated) it seems quite possible for this
relatively small population of clusters to have been formed from objects 
like the lower mass molecular complexes seen in the Antennae.

Young massive star clusters have been found even in relatively low-luminosity
dwarf irregular galaxies \citep*{ghg01,h00,b02}. \citet{b02} note that,
although it is rare for a dwarf galaxy to form luminous star clusters,
when they do, they tend to form several clusters with similar ages. This 
result suggests that star cluster formation is concentrated to a localized
region. \citet{b02} emphasize that a lack of shear is probably the
most important difference between dwarf irregular and spiral galaxies,
and suggest that triggered large-scale flows, possibly by an interaction,
or ambient instabilities in a shear-free environment can make the clouds
that form young massive star clusters in dwarf galaxies. 
%**vv
Given the radically
disturbed velocity fields seen in the Antennae, reduced shear might also
be a factor in cloud formation in this system. An important difference
between the Antennae and dwarf galaxies is the sheer size of the gas
reservoir \citep{g01}; perhaps combining a large gas reservoir with
localized regions of
reduced shear is the key to explaining the explosion
of star cluster formation within the Antennae.
%**^^

\section{Conclusions \label{concl}}

We have used sensitive CO J=1-0 observations of the Antennae to
study the detailed properties of the molecular clouds 
with the goal of understanding the
prodigious formation of young massive star clusters in this nearby
merger system. We have identified a total of $\sim 100$ 
clouds in the data cube with masses ranging from $2\times 10^6$ 
to $9\times 10^8$ M$_\odot$.
This sample of extragalactic molecular clouds is unique in the
total number of clouds identified and in the mass range probed by the
observations.

Above our 5$\sigma$ completeness limit of $5\times 10^6$ M$_\odot$,
the cloud mass function has a slope of $-1.4 \pm 0.1$. This mass
function slope is very similar to that seen in molecular clouds and
molecular cloud cores in the Galaxy, and is somewhat steeper than
the mass function slope of -2 estimated for the luminous young star clusters
in the Antennae by \citet{zf99}. Our data suggest that the molecular
interstellar medium in galaxies is governed by the same mass function
slope over 8 to 9 orders of magnitude in mass, from 1-10 M$_\odot$
up to $10^9$ M$_\odot$.

We have compared the Antennae with the nearby spiral galaxy M51,
for which similar sensitivity and spatial resolution CO 
observations exist. Although the
two galaxies have similar gas surface densities and total gas masses,
the molecular clouds in M51 are an order of magnitude less massive
than those in the Antennae. In addition, M51 has a much smaller
population of young massive star clusters, perhaps 50 times smaller
than that of the Antennae. One significant difference between the
gas in the two galaxies is that M51 exhibits a smooth velocity
field, while the velocity field in the Antennae is highly disordered.
%** no changes here - aug 27,2003
One possibility is that young massive star cluster formation in M51 is
suppressed by its inability to form very massive gas clouds due
to high shear in its differentially rotating disk.

Comparing the CO data with the locations of the youngest clusters shows
that many of these clusters lie in the CO-rich overlap region. The
extremely high extinction ($ A_v \sim 100$ mag) towards the CO peaks 
means that the star clusters must lie in front of most of the molecular
gas in this region. Interestingly, some of these youngest clusters are
found as much as 2 kpc from regions with detectable molecular gas. This
result implies that either some young massive star clusters can form from
clouds less massive than $5\times 10^6$ M$_\odot$, or these clusters
have already destroyed their parent molecular clouds, or both.

We have combined our CO data with published radio and optical data to
sketch out the recent star formation history of the Antennae and to
speculate on how star formation will proceed in the future. The
relative numbers of very young massive 
star clusters and thermal and non-thermal
radio continuum sources are consistent with a constant star formation
rate over the last 10-20 Myr. If the star formation rate has been
constant over the last 100 Myr, then the relative numbers of
star clusters with ages less than 10 Myr and 20-100 Myr 
implies that a very large fraction of the
star clusters formed must evaporate or be destroyed
\citep{zf99}. On the other hand, if most star clusters 
survive for long periods, we would have to be witnessing an
extremely short-duration starburst in order for the final
specific frequency of the Antennae not to exceed 
the typical specific frequency seen in elliptical galaxies
by a wide margin.

The abundant supply of molecular gas
seen in the extremely massive molecular clouds suggests that star
cluster formation could easily proceed for at least a crossing time
of one of these large clouds, or about 30 Myr. Indeed, the large
reservoir of molecular gas measured by \citet{g01} suggests that star
cluster formation could continue for quite some time. The current
distribution of gas and stars in the Antennae suggests a mode of
star cluster formation where portions of the galactic disks are lit
up with recently formed clusters (for example, regions like the
north-western arc) while other regions are forming the next generation
of star clusters (currently in the overlap region).

We have compared our observations with two different models to
explain the formation of young massive star clusters. 
The identification of extremely
massive gas clouds in the Antennae means that the model of \citet{hp94},
which envisions globular cluster or super star cluster formation
as a scaled up version of Galactic star cluster formation,
is a potentially viable model. The model advanced by \citet{s96},
which requires a source of high-pressure in order to collapse
pre-existing giant molecular clouds to form stars with a much
higher efficiency, may also be viable given recent constraints
on the pressure in the Antennae from X-ray observations
\citep{f03}. 
In addition, higher pressures can also be a feature of the
\citet{hp94} model, so that pressure alone cannot distinguish
between the models. Understanding the formation of super star
clusters in dwarf galaxies, where there may not be a global
source of enhanced pressure, may provide additional interesting
constraints on this question. Ultimately, observations with new 
telescopes such as the Atacama Large Millimeter Array, 
which will have sufficient resolution
and sensitivity to probe the small physical scales on which
young massive star clusters form, will probably be required to fully understand
this intriguing mode of star formation.

\acknowledgments 
We thank the referee for useful comments which improved the discussion,
particularly in \S5 and 6. The research of CDW
is supported through grants from the Natural
Sciences and Engineering Research Council of Canada.  
VC would like to acknowledge the support of JPL contract 960803.
The Owens Valley Millimeter Array is operated by the California
Institute of Technology and is supported by
NSF grant AST96-13717. CDW acknowledges the hospitality of
the Aspen Center for Physics and useful conversations with
Susan Neff and Jim Ulvestad.

\appendix
\section{Properties of Individual Super Giant Molecular Complexes
in the Antennae}
\label{appendixA}

The properties of the individual super giant molecular
complexes (such as
position, flux, and velocity) are given in Table~\ref{tbl-all}. The clouds were
identified using  the program
CLFIND with a contour level of 0.11 Jy beam$^{-1}$ ($2\sigma$) and
their properties were measured using the program CLSTATS
\citep{w94}. 
The 5$\sigma$ mass sensitivity limit of
the data set is $5\times 10^6$ M$_\odot$.

\section{The CO-to-H$_2$ Conversion Factor in the Antennae}
\label{appendixB}

Most of the clouds identified in the Antennae are unresolved, and so
their masses can only be calculated from their CO flux. It is important,
therefore, to estimate the CO-to-H$_2$ conversion factor that is
appropriate for the Antennae using the few clouds that are large
enough that their true diameters can be deconvolved from the
synthesized beam. We excluded
clouds in the two galactic nuclei or near bright mid-infrared peaks
from this analysis because the conversion factor
could be modified in these regions due to intense star formation
or high pressure \citep{b96,sdrb97}.

We analyzed the data cube with CLFIND and CLSTATS using
three different contour levels (0.10, 0.11=2$\sigma$, and 0.12 Jy beam$^{-1}$) 
as described above.  The integrated intensity map for
each cloud that appeared resolved in both dimensions from the output
from CLSTATS was inspected to see if it was in fact resolved. (Clouds
may appear resolved from their sizes but not actually be resolved if,
for example, the cloud is actually made up of two peaks, or if the
cloud is highly elongated in one direction.) We identified four, three, and
two clouds that appeared to be resolved when identified with contour
levels of 0.10, 0.11, and 0.12 Jy beam$^{-1}$, respectively
(Table \ref{tbl-XCO}). Only one
cloud was found in common to both the 0.10 and 0.11 Jy beam$^{-1}$ processing.
The remaining seven clouds were identified when contoured at different
levels, but, due to small changes in how flux was assigned to each
cloud, they did not always appear to be resolved spatially.

For each resolved cloud, the radius and velocity width were measured
using CLSTATS while the total CO flux was measured from a gain-corrected
zeroth moment map of each individual cloud. We used these measurements to
calculate both the virial mass and
the molecular mass. The virial mass is given by
$$M_{vir} = 198 \Delta V_{FWHM}^2 R_{pc} \hskip3pt \rm{M_\odot}$$
where $\Delta V_{FWHM}$ is the full width half maximum velocity in
km s$^{-1}$ and $R_{pc} = \sqrt{Area/\pi}$ is the deconvolved radius 
of the cloud in parsecs. The molecular mass is given by
$$M_{mol} = 1.61 \times 10^4 D_{Mpc}^2 S_{CO} \hskip3pt \rm{M_\odot}$$
where $D_{Mpc}$ is the distance to the cloud in Mpc and $S_{CO}$ is
the CO integrated intensity in Jy km s$^{-1}$ \citep{ws90}.
The ratio of the assumed value for CO-to-H$_2$ conversion factor in
the Galaxy relative to the true value in the Antennae 
can then be determined from
the ratio of the molecular mass to the virial mass. The average value
of this ratio for the eight resolved clouds from the three different
identification runs is $1.3\pm0.3$ with a standard deviation of 0.7.
This result suggests that the CO-to-H$_2$ conversion factor may be
slightly smaller in the Antennae than in the Milky Way. However,
given that the ratio is consistent with equal conversion factors
within the uncertainties, we have chosen to use the standard Galactic
value ($3 \times 10^{20}$ H$_2$ cm$^{-2}$ (K km s$^{-1}$)$^{-1}$)
for ease of comparison with other work.

Clouds that were resolved but were not used to determine the
CO-to-H$_2$ conversion factor due
to their location in the nuclei or the mid-infrared bright part of the 
 overlap region are given in Table \ref{tbl-res}. Cross-identifications
with \citet{w00} are also given; note that, in some cases, several
complexes in Table \ref{tbl-all} and Table \ref{tbl-res} correspond
to a single super giant molecular complex from \citet{w00}. A comparison
of the virial and molecular masses in Table \ref{tbl-res} suggests
that the CO-to-H$_2$ factor may be smaller than the adopted value in
several of these clouds. In particular, the complexes in the nucleus
of NGC 4038 all have molecular masses that exceed their virial masses
by at least a factor of three. This result suggests that the CO emission is
over-luminous in this galactic nucleus, perhaps similar
to the effect seen in ultraluminous infrared galaxies
\citep{sdrb97}. We will discuss the properties of the nuclear regions
in a future paper (Wilson, Madden, \& Charmandaris, in preparation).

\clearpage

\begin{deluxetable}{lccclc}
\tabletypesize{\scriptsize}
\tablecaption{Massive Molecular Complexes in Other Galaxies
\label{tbl-other-gmas}}
\tablewidth{0pt}
\tablehead{
\colhead{Galaxy} & \colhead{Beam Size} & 
\colhead{Distance}  
& \colhead{Maximum Mass}  & \colhead{Notes} & 
\colhead{References\tablenotemark{a}}     
\\
 & \colhead{(pc)} & 
\colhead{(Mpc)}  &
\colhead{($10^8$ M$_\odot$)}  & 
\colhead{}     & \colhead{}
}
\startdata
M83 & $110\times  55$ & 3.3 &  0.1 & 10 complexes & 1,10 \\ %OK
NGC 5055 & $220 \times 170$ & 7.2 &  0.5 & 17 complexes & 2,11 \\ %OK
M51 & $370\times 290$ & 8.6 &  0.5 & 26 complexes & 3,12 \\
NGC 1068 & $210\times 210$ & 14.4 & 4.5 & 38 complexes & 4,11 \\ %OK
M100 & $360\times 280$ & 16 &  1.2 & 7 complexes & 5,13 \\ %OK
NGC 4414 & $480\times 480$ & 19 &  0.4 & 8 complexes & 6,14 \\ %OK
NGC 4038/39 & $480 \times 310$ & 19 & 8.9 & 86 complexes & 7,15 \\ %OK
Arp 299 & $750\times 510$ & 42 & 40 & nucleus of IC694 & 8,15 \\ %OK
Arp 220 & $210\times 190$ & 77 & 10 & nucleus, $X_{CO}$ uncertain & 9,15 \\ %OK
\enddata
\tablecomments{Masses have been scaled to the same CO-to-H$_2$
conversion factor used in this paper, except for Arp 220 where
a value of 2.5 times smaller was used.}
\tablenotetext{a}{References for molecular data and distance.
1. \citet{r99} 2. \citet{t97} 3. \citet{rk90} 4. \citet{p91}
5. \citet{r95} 6. \citet{sa96} 7. this paper 8. \citet{c99}
9. \citet{s99} 10. Distance to group member NGC 5253 from \citet{g00} 
11. \citet{tf88} 12. \citet{f97} 13. \citet{f96} 14. \citet{t98}
15. $H_o = 75$ km s$^{-1}$ Mpc$^{-1}$.}

\end{deluxetable}

\newpage

\begin{deluxetable}{lccccccc}
\tabletypesize{\scriptsize}
\tablecaption{Giant Molecular Complexes in the Antennae \label{tbl-all}}
\tablewidth{0pt}
\tablehead{
\colhead{ID\tablenotemark{a}} & \colhead{$\alpha$} & 
\colhead{$\delta$}  
& \colhead{$S_{CO}$}  & \colhead{$T_B$\tablenotemark{b}} &
\colhead{$V_{lsr}$}     & \colhead{$\Delta V_{FWHM}$}  &
\colhead{$M_{mol}$} 
\\
 & \colhead{(2000)} & 
\colhead{(2000)}  &
\colhead{(Jy km s$^{-1}$)}  & 
\colhead{(K)}     & \colhead{(km s$^{-1}$)}
& \colhead{(km s$^{-1}$)} &\colhead{($10^7$ M$_\odot$)}   
}
\startdata
 20 & 12:01:53.1 & -18:53:15 &  16.2 & 3.1 & 1683 & 21 &   9.4 \\ 
 86 & 12:01:53.1 & -18:53:15 &   1.8 & 2.6 & 1755 &  6 &   1.0 \\ 
 22\tablenotemark{c} & 12:01:53.1 & -18:53:14 &  23.2 & 4.0 & 1703 & 37 &  13.5 \\ 
 58\tablenotemark{c} & 12:01:53.2 & -18:53:13 &  47.4 & 2.8 & 1610 & 56 &  27.5 \\ 
 72 & 12:01:55.1 & -18:53:10 &   0.4 & 2.1 & 1641 &  5 &   0.2 \\ 
 32 & 12:01:54.2 & -18:53:08 &   2.9 & 1.9 & 1464 & 16 &   1.7 \\ 
 79 & 12:01:53.7 & -18:53:08 &   5.5 & 2.0 & 1683 & 27 &   3.2 \\ 
 65 & 12:01:53.7 & -18:53:07 &   0.6 & 1.6 & 1620 &  9 &   0.3 \\ 
  7\tablenotemark{c} & 12:01:54.9 & -18:53:06 &  11.2 & 3.5 & 1501 & 11 &   6.5 \\ 
 80 & 12:01:53.4 & -18:53:05 &   0.9 & 1.9 & 1693 &  9 &   0.5 \\ 
 44 & 12:01:52.4 & -18:53:04 &   1.4 & 2.5 & 1558 &  7 &   0.8 \\ 
 69 & 12:01:55.9 & -18:53:04 &   1.3 & 2.8 & 1631 &  6 &   0.7 \\ 
 13 & 12:01:54.9 & -18:53:03 &   2.1 & 2.3 & 1480 &  5 &   1.2 \\ 
 12 & 12:01:54.6 & -18:53:02 &  40.6 & 2.8 & 1485 & 53 &  23.6 \\ 
  8\tablenotemark{c} & 12:01:55.0 & -18:53:01 &  42.6 & 3.8 & 1532 & 31 &  24.8 \\ 
 15\tablenotemark{c} & 12:01:54.9 & -18:53:01 &  22.7 & 2.6 & 1568 & 49 &  13.2 \\ 
  9\tablenotemark{c} & 12:01:54.9 & -18:53:01 &  11.3 & 3.7 & 1553 & 12 &   6.6 \\ 
 66 & 12:01:54.8 & -18:52:59 &   4.7 & 1.8 & 1620 & 22 &   2.7 \\ 
 34 & 12:01:55.4 & -18:52:58 &   5.4 & 2.2 & 1537 & 15 &   3.1 \\ 
 38 & 12:01:55.8 & -18:52:58 &   0.4 & 2.3 & 1542 &  6 &   0.2 \\ 
 77 & 12:01:54.7 & -18:52:56 &   2.5 & 1.5 & 1667 & 12 &   1.5 \\ 
 54 & 12:01:54.9 & -18:52:55 &   1.4 & 2.0 & 1584 &  7 &   0.8 \\ 
 63 & 12:01:54.9 & -18:52:55 &   0.4 & 1.5 & 1605 &  7 &   0.2 \\ 
 73 & 12:01:54.7 & -18:52:55 &   1.8 & 1.5 & 1641 & 18 &   1.0 \\ 
 14 & 12:01:54.8 & -18:52:54 &   4.3 & 2.1 & 1521 & 55 &   2.5 \\ 
 40 & 12:01:55.1 & -18:52:54 &   0.3 & 1.6 & 1547 &  5 &   0.2 \\ 
 33 & 12:01:55.4 & -18:52:54 &   1.6 & 1.9 & 1475 &  5 &   0.9 \\ 
 59 & 12:01:54.8 & -18:52:53 &   2.4 & 2.0 & 1594 & 17 &   1.4 \\ 
 70 & 12:01:54.7 & -18:52:53 &   0.4 & 1.4 & 1631 &  6 &   0.2 \\ 
  4\tablenotemark{c} & 12:01:54.8 & -18:52:52 &  23.8 & 2.8 & 1449 & 23 &  13.8 \\ 
  5\tablenotemark{c} & 12:01:54.9 & -18:52:52 &  44.0 & 3.4 & 1480 & 33 &  25.6 \\ 
 47 & 12:01:54.7 & -18:52:52 &   2.3 & 1.4 & 1563 & 13 &   1.3 \\ 
 37 & 12:01:54.7 & -18:52:52 &   0.9 & 1.5 & 1537 & 10 &   0.5 \\ 
 11 & 12:01:54.7 & -18:52:50 &   8.6 & 2.2 & 1417 & 23 &   5.0 \\ 
  6 & 12:01:55.4 & -18:52:49 & 109.0 & 4.1 & 1501 & 39 &  63.3 \\ 
 25 & 12:01:54.8 & -18:52:49 &   3.5 & 1.6 & 1381 & 13 &   2.1 \\ 
 26 & 12:01:55.4 & -18:52:49 &   1.4 & 1.8 & 1391 & 16 &   0.8 \\ 
 27 & 12:01:54.7 & -18:52:49 &   1.4 & 1.4 & 1397 & 15 &   0.8 \\ 
 29 & 12:01:55.4 & -18:52:48 &   6.1 & 1.9 & 1428 & 15 &   3.6 \\ 
 31 & 12:01:55.4 & -18:52:48 &   3.6 & 2.0 & 1459 &  8 &   2.1 \\ 
 24 & 12:01:55.4 & -18:52:48 &   1.5 & 2.0 & 1381 &  8 &   0.9 \\ 
 23 & 12:01:55.4 & -18:52:47 &   8.3 & 2.3 & 1371 & 22 &   4.8 \\ 
 30 & 12:01:55.4 & -18:52:47 &   2.2 & 2.0 & 1443 & 11 &   1.3 \\ 
 83 & 12:01:55.4 & -18:52:47 &   1.4 & 1.8 & 1709 &  6 &   0.8 \\ 
 28 & 12:01:55.5 & -18:52:46 &   3.4 & 2.0 & 1402 & 10 &   2.0 \\ 
 10 & 12:01:54.7 & -18:52:44 &  14.9 & 2.8 & 1589 & 23 &   8.6 \\ 
 67 & 12:01:54.4 & -18:52:44 &   5.7 & 1.8 & 1620 & 13 &   3.3 \\ 
 74 & 12:01:54.5 & -18:52:44 &   0.6 & 1.4 & 1646 &  5 &   0.3 \\ 
 16 & 12:01:55.1 & -18:52:41 &  16.0 & 2.5 & 1568 & 21 &   9.3 \\ 
 17 & 12:01:55.1 & -18:52:40 &  11.4 & 3.1 & 1584 & 14 &   6.6 \\ 
 18 & 12:01:55.1 & -18:52:38 &   7.3 & 2.9 & 1599 & 13 &   4.3 \\ 
 19 & 12:01:54.7 & -18:52:34 &  23.6 & 2.2 & 1620 & 26 &  13.7 \\ 
 35 & 12:01:53.2 & -18:52:33 &   0.6 & 1.5 & 1527 &  5 &   0.4 \\ 
 84 & 12:01:55.3 & -18:52:33 &   1.9 & 1.8 & 1719 &  6 &   1.1 \\ 
 53 & 12:01:53.0 & -18:52:32 &   1.6 & 1.7 & 1579 &  7 &   0.9 \\ 
 76 & 12:01:51.3 & -18:52:31 &   2.1 & 2.5 & 1651 &  6 &   1.2 \\ 
 75 & 12:01:54.4 & -18:52:30 &   2.0 & 1.4 & 1646 & 17 &   1.2 \\ 
 51 & 12:01:54.4 & -18:52:29 &   9.2 & 1.6 & 1568 & 44 &   5.4 \\ 
 55 & 12:01:55.1 & -18:52:29 &   4.7 & 2.1 & 1589 &  7 &   2.7 \\ 
 42 & 12:01:53.5 & -18:52:28 &   0.8 & 1.4 & 1553 &  5 &   0.5 \\ 
 68 & 12:01:54.5 & -18:52:28 &   3.5 & 1.5 & 1625 & 16 &   2.1 \\ 
 52 & 12:01:55.1 & -18:52:26 &   2.2 & 1.9 & 1568 &  8 &   1.3 \\ 
 60 & 12:01:55.9 & -18:52:25 &   1.7 & 2.4 & 1594 &  6 &   1.0 \\ 
 71 & 12:01:50.5 & -18:52:24 &  16.5 & 1.9 & 1667 & 41 &   9.6 \\ 
 45 & 12:01:51.9 & -18:52:22 &   0.7 & 1.6 & 1558 &  5 &   0.4 \\ 
 78 & 12:01:52.0 & -18:52:20 &   4.1 & 1.6 & 1677 & 13 &   2.4 \\ 
 56 & 12:01:53.3 & -18:52:16 &   2.1 & 1.4 & 1584 & 22 &   1.2 \\ 
 48 & 12:01:53.7 & -18:52:12 &   1.3 & 1.4 & 1563 &  6 &   0.7 \\ 
 64 & 12:01:52.1 & -18:52:08 &   4.7 & 1.3 & 1615 & 28 &   2.7 \\ 
 62 & 12:01:50.9 & -18:52:07 &   5.4 & 1.6 & 1599 & 19 &   3.2 \\ 
 57 & 12:01:50.8 & -18:52:06 &   2.4 & 1.4 & 1589 & 12 &   1.4 \\ 
 61 & 12:01:50.5 & -18:52:06 &   0.6 & 1.5 & 1594 &  5 &   0.4 \\ 
 49 & 12:01:53.1 & -18:52:05 &   1.5 & 1.6 & 1563 &  8 &   0.9 \\ 
 81 & 12:01:53.8 & -18:52:04 &   1.7 & 1.4 & 1693 &  9 &   1.0 \\ 
  1 & 12:01:53.0 & -18:52:02 & 152.4 & 6.7 & 1620 & 45 &  88.6 \\ 
  2 & 12:01:53.0 & -18:52:01 &  50.8 & 5.8 & 1651 & 16 &  29.5 \\ 
  3 & 12:01:53.0 & -18:52:01 &  35.9 & 4.5 & 1667 & 18 &  20.9 \\ 
 21 & 12:01:53.0 & -18:52:00 &  19.5 & 2.4 & 1688 & 27 &  11.3 \\ 
 50 & 12:01:51.8 & -18:51:58 &   1.2 & 1.3 & 1563 &  8 &   0.7 \\ 
 39 & 12:01:50.7 & -18:51:57 &   9.1 & 1.7 & 1547 & 39 &   5.3 \\ 
 82 & 12:01:52.3 & -18:51:57 &   4.7 & 1.5 & 1703 & 20 &   2.7 \\ 
 36 & 12:01:50.7 & -18:51:54 &   7.1 & 1.5 & 1532 & 25 &   4.1 \\ 
 85 & 12:01:52.3 & -18:51:54 &   1.9 & 1.6 & 1729 &  7 &   1.1 \\ 
 41 & 12:01:51.1 & -18:51:52 &  17.2 & 2.1 & 1568 & 29 &  10.0 \\ 
 46 & 12:01:51.8 & -18:51:41 &   4.5 & 2.3 & 1558 & 11 &   2.6 \\ 
 43 & 12:01:51.8 & -18:51:38 &   4.9 & 2.5 & 1553 & 17 &   2.8 \\ 
\enddata
\tablecomments{A distance to the Antennae of 19 Mpc is assumed throughout.}
\tablenotetext{a}{Cloud ID number assigned by CLFIND algorithm.}
\tablenotetext{b}{Observed peak brightness temperature excess above the 
2.74 K cosmic background.}
\tablenotetext{c}{These clouds belong to the super giant molecular
complexes discussed in \citet{w00}: complexes 4, 5: SGMC 2; complexes
7, 8, 9: SGMC 4; complex 15: SGMC 5; complexes 22, 58: part of NGC 4039.}

\end{deluxetable}

\begin{deluxetable}{lccccccccc}
\tabletypesize{\scriptsize}
\tablecaption{Resolved Clouds Used to Determine $X_{CO}$ \label{tbl-XCO}}
\tablewidth{0pt}
\tablehead{
\colhead{ID} 
& \colhead{$S_{CO}$}  & 
\colhead{$R$}     & \colhead{$\Delta V_{FWHM}$} &\colhead{$M_{vir}$}    &
\colhead{$M_{mol}$} & \colhead{CLFIND contour} 
\\
 & 
\colhead{(Jy km s$^{-1}$)}  & 
\colhead{(pc)}     & \colhead{(km s$^{-1}$)} &\colhead{(10$^8$ M$_\odot$)}    &
\colhead{(10$^8$ M$_\odot$)} & \colhead{(Jy beam$^{-1}$)}
}
\startdata
17 & %12:01:55.1 & -18:52:40 & 
28.6 & 580 & 33 & 1.3   & 1.7 & 0.12 \\ %cloud 13 in 0.12 nomenclature
39 & %12:01:50.7 & -18:51:58 & 
\pad 9.6 & 430 & 23 & \pad 0.45 & \pad 0.56 & 0.12 \\ % cloud 28 in 0.12 nomenclature
19\tablenotemark{a} & %12:01:54.7 & -18:52:34 & 
23.6 & 560 & 24 & \pad 0.64 & 1.4 & 0.11 \\ 
36 & %12:01:50.7 & -18:51:54 & 
\pad 7.1 & 430 & 25 & \pad 0.53 & \pad 0.41 & 0.11 \\
71 & %12:01:50.8 & -18:52:13 & 
16.5 & 610 & 41 & 2.0  & \pad 0.96 & 0.11 \\
41 & %12:01:51.1 & -18:51:52 & 
16.9 & 610 & 20 & \pad 0.48 & \pad 0.98 & 0.10 \\ % cloud 15 in 0.10 nomenclature
62 & %12:01:50.9 & -18:52:07 & 
19.3 & 600 & 46 & 2.5  & 1.1 & 0.10 \\ % cloud 75 in 0.10 nomenclature
46 & %12:01:51.8 & -18:51:40 & 
\pad 5.7 & 410 & 14 & \pad 0.16 & \pad 0.33 & 0.10 \\ % cloud 44 in 0.10 nomenclature
\enddata
\tablecomments{A distance to the Antennae of 19 Mpc is assumed throughout.
Values for $S_{CO}$ and $\Delta V_{FWHM}$ may differ from Table \ref{tbl-all}
for the clouds identified with a different contour level.}
\tablenotetext{a}{This cloud was resolved for both the 0.11 and 0.10 
Jy beam$^{-1}$ contour levels. The values for all parameters are
the average of two values.}

\end{deluxetable}

\begin{deluxetable}{lccccl}
\tabletypesize{\scriptsize}
\tablecaption{Other Resolved Clouds \label{tbl-res}}
\tablewidth{0pt}
\tablehead{
\colhead{ID} & \colhead{$R$}     & \colhead{$\Delta V_{FWHM}$} 
&\colhead{$M_{vir}$}    & \colhead{$M_{mol}$} & \colhead{ID from \citet{w00}} 
\\
 & \colhead{(pc)}     & \colhead{(km s$^{-1}$)} 
&\colhead{(10$^8$ M$_\odot$)}    &
\colhead{(10$^8$ M$_\odot$)} & 
}
\startdata
  1 & 760 & 45 &  3.1 &  8.9 & NGC 4038 \\ 
  2 & 600 & 16 &  0.3 &  3.0 & NGC 4038 \\ 
  3 & 630 & 18 &  0.4 &  2.1 & NGC 4038 \\ 
  6 & 670 & 39 &  2.0 &  6.3 & SGMC 1 \\ 
 12 & 670 & 53 &  3.7 &  2.4 & SGMC 3 \\ 
 20 & 440 & 21 &  0.4 &  0.9 & NGC 4039 \\ 
 21 & 480 & 27 &  0.7 &  1.1 & NGC 4038 \\ 
\enddata

\end{deluxetable}

\newpage

\clearpage

\figcaption[figure1.ps]{(a) Integrated CO intensity map from \citet{w00}.
Contour levels are 1,2,4,6,8,10,15,20,25...50,60...,90 Jy beam$^{-1}$
km s$^{-1}$. This map was made using CLPLOT and including only 
molecular complexes with velocity widths greater than or equal to 15.6
km s$^{-1}$ found with CLFIND. This image has been corrected for the 
fall-off in sensitivity due to the primary beam.
(b) Integrated intensity map made using the same data set and criteria, 
but including the clouds that were missed due to a bug in the CLFIND
program.
(c) A standard zeroth moment map made from the same data cube used to
identify molecular complexes. The 50\% gain limit of the telescope primary
beam is clearly visible in this figure.
Negative bowls exist in this map (see text) but are not shown for
ease of comparison with the other two maps.
\label{fig1}}
%(a) (Fig 1): all_clouds_normg_2000_frame
%(b) moment0cl_gain in /1/wilson/NGC4038/Compare_images
%(c) moment0_gain  in /1/wilson/NGC4038/Compare_images

\figcaption[figure4a.ps]{First moment map of the CO emission in the Antennae.
The color scale runs from 1460 km s$^{-1}$ to 1660 km s$^{-1}$, while
the contour levels run from 1460 to 1660 km s$^{-1}$ in steps of
20 km s$^{-1}$. Note in particular the disturbed velocity field in the
overlap region.
\label{mom1}}

\figcaption[figure2.ps]{The differential mass function for the molecular
complexes in the Antennae. Mass functions are shown for three identification
runs with different contour levels; each identification run is also shown
for two different binnings of the mass function separated by 0.1 in $\log(M)$.
The 5$\sigma$ completeness limit is indicated by the dashed line.
The contour level in Jy beam$^{-1}$ and the slope derived from a least squares
fit to the points above the completeness limit are given in each panel.
The average slope of $-1.4\pm0.1$ is very similar to the slope observed
for Giant Molecular Clouds ($10^3-10^6$ M$_\odot$) in the Milky Way.
\label{fig-diff-mass}}

\figcaption[figure7_new.ps]{Locations of young, red super star clusters 
using the coordinates from
\citet{z01} overlaid on the CO contours from Figure 1b. Roughly
40\% of these clusters are found in the overlap region, which
covers only 10\% of the area of the total CO map. However,
some young clusters are found as far as 1-2 kpc from strong CO
emission, which suggests they either formed from a lower mass molecular
complex or they have already destroyed their parent cloud. The 
locations of the eight resolved clouds from  Table 
\ref{tbl-XCO} are also indicated.
\label{optcoplot}}

% figures are in /1/wilson/NGC4038/ various subdirectories

\clearpage

\epsscale{0.50}
\plotone{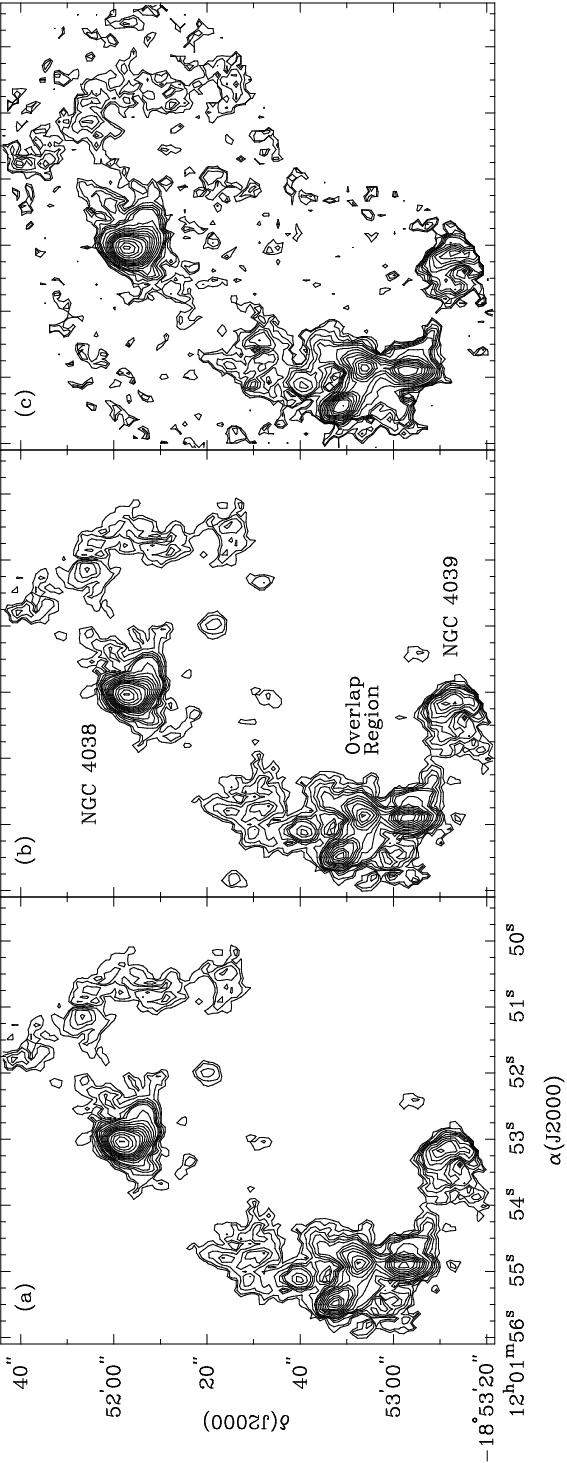}
\vfil
\epsscale{1.0}
\plotone{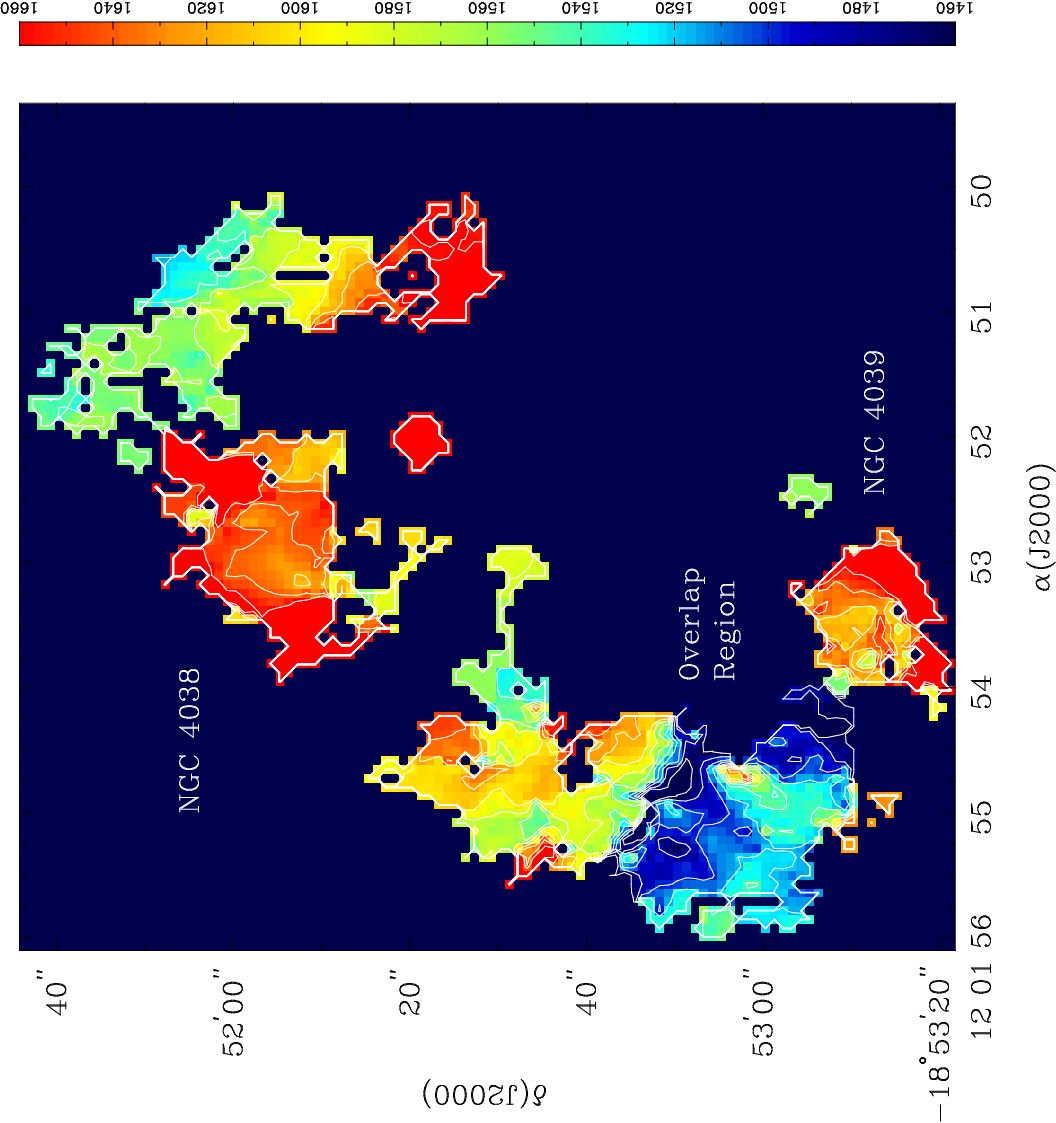}
\vfil
\plotone{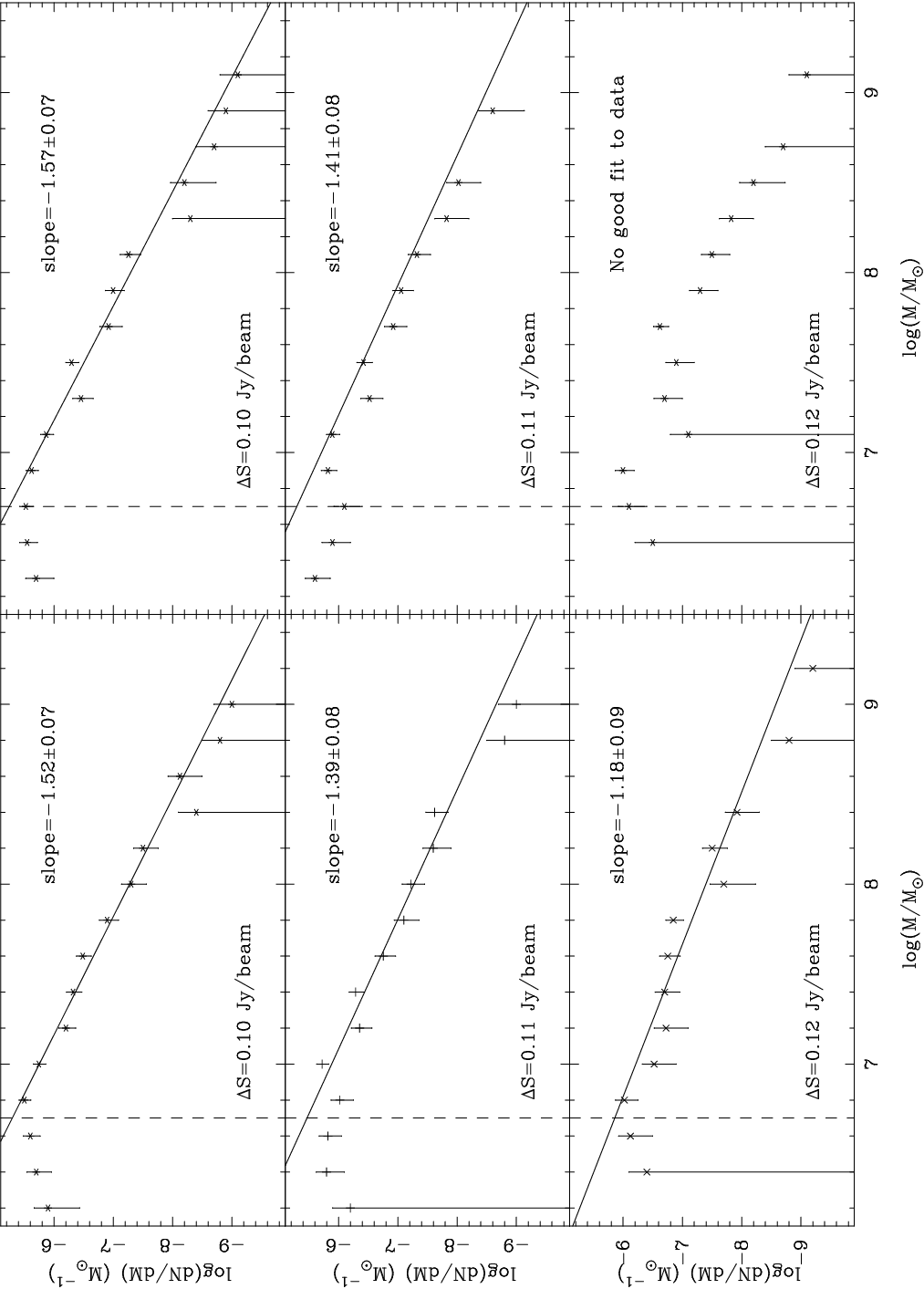}
\vfil
\plotone{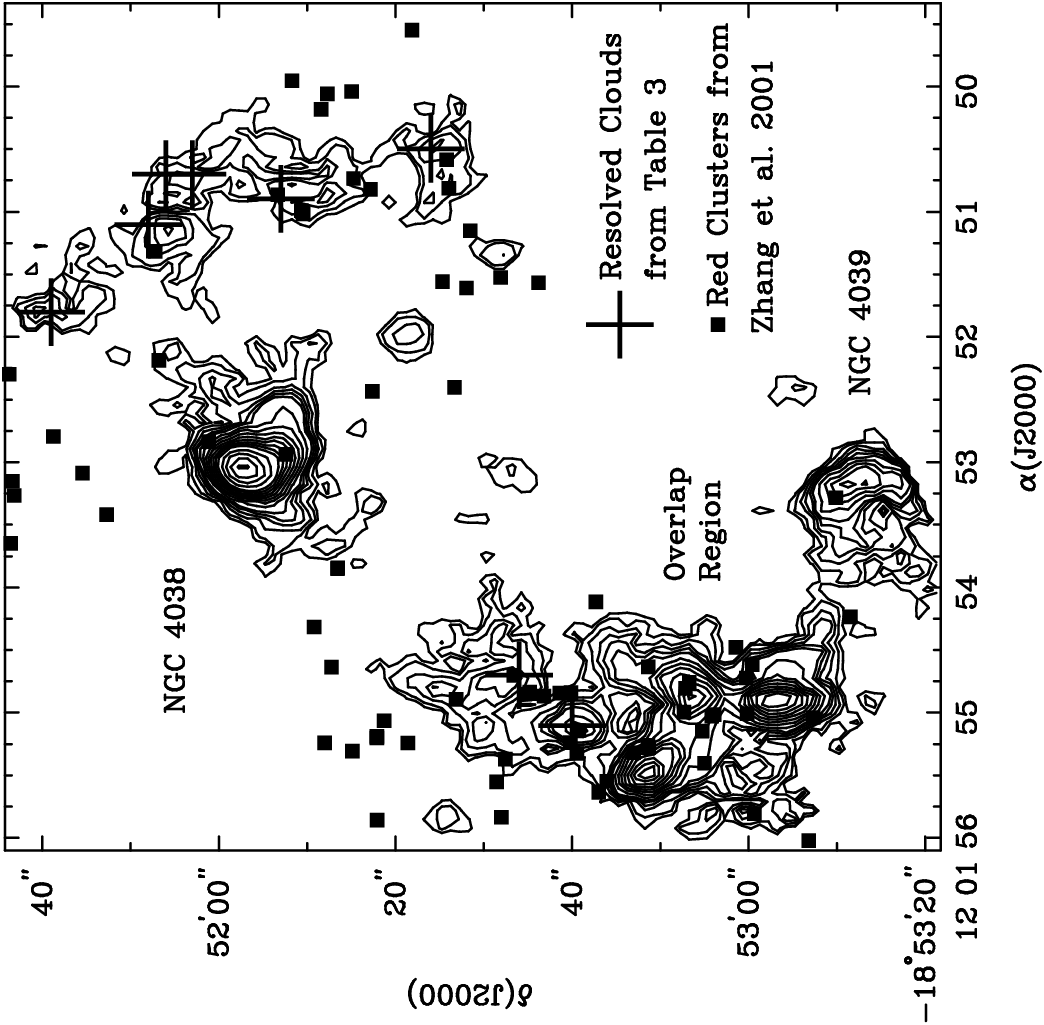}


\newpage
\begin{thebibliography}{}

\bibitem[Aalto et al.(1999)]{a99}Aalto, S., H\"uttemeister, S.,
Scoville, N. Z. \& Thaddeus, P., 1999, \apj, 522, 165

\bibitem[Billett et al.(2002)Billett, Hunter \& Elmegreen]{b02} 
Billett, O. H., Hunter, D. A., \& Elmegreen, B. G., 2002, \aj, 123, 1454

\bibitem[Bryant \& Scoville(1996)]{b96}Bryant, P. M., \&  Scoville, N. Z.,
1996, \apj, 457, 678

\bibitem[Casoli et al.(1999)]{c99}Casoli, F.,  Willaime, M.-C., Viallefond, F.,
 \& Gerin, M., 1999, \aap, 346, 663

\bibitem[Cen(2001)]{c01}Cen, R., 2001, \apj, 560, 592 

\bibitem[Cohen et al.(1988)]{c88}
Cohen, R. S., Dame, T. M., Garay, G., Montani, J., Rubio, M.,
\& Thaddeus, P., 1998, \apj, 331, L95

\bibitem[Elmegreen(1989)]{e89}Elmegreen, B. G., 1989, \apj, 338, 178

\bibitem[Elmegreen \& Efremov(1997)]{ee97}Elmegreen, B. G., \& Efremov, Y. N.,
1997, \apj, 480, 235

\bibitem[Evans \& Lada(1991)]{el91} Evans, N. J., \& Lada, E. A., 1991,
  in Fragmentation of Molecular Clouds and Star Formation (IAU
  Symposium 147), eds. E. Falgarone, F. Boulanger, \& G. Duvert, 
  [Kluwer: Dordrecht], 293

\bibitem[Fabbiano et al.(2003)]{f03} Fabbiano, G., Krauss, M., Zezas, A., Rots,
A. \& Neff, S., 2003, \apj, in press

\bibitem[Fall \& Rees(1985)]{fr85} Fall, S. M., \& Rees, M. J., 1985, \apj, 
  298, 18

\bibitem[Fall \& Zhang(2001)]{fz01}  Fall, S. M., \& Zhang, Q.,  2001, 
\apj, 561, 751

\bibitem[Feldmeier et al.(1997)Feldmeier, Ciardullo, \& Jacoby]{f97} Feldmeier,
J. J., Ciardullo, R., \& Jacoby, G. H., 1997, \apj, 479, 231

\bibitem[Ferrarese et al.(1996)]{f96} Ferrarese, L., et al., \apj, 464, 568

\bibitem[Ferrarese et al.(2000)]{f00} Ferrarese, L., et al., \apj, 529, 745

\bibitem[Gao et al.(2001)]{g01} Gao, Y., Lo, K. Y., Lee, S.-W., \& Lee, T.-H.,
  2001, \apj, 548, 172

\bibitem[Garcia-Burillo et al.(1993)]{g93} Garcia-Burillo, S., Gu\'elin,
M., \& Cernicharo, J., 1993, \aap, 274, 123

\bibitem[Gelatt et al.(2001)Gelatt, Hunter, \& Gallagher]{ghg01}
Gelatt, A. E., Hunter, D. A., \& Gallagher, J. S., 2001, \pasp, 113, 142

\bibitem[Gibson et al.(2000)]{g00}Gibson, B. K., et al., 2000, \apj,
529, 723

\bibitem[Harris(2001)]{harris01} Harris, W. E., 2001, in Star Clusters,
eds. L. Labhardt \& B. Bingelli, [Springer-Verlag: Berlin], 223

\bibitem[Harris \& Pudritz(1994)]{hp94} Harris, W. E., \& Pudritz, R. E.,
  1994, \apj, 429, 177

\bibitem[Heyer et al.(2001)]{heyer01} Heyer, M. H., Carpenter, J. M., \&
Snell, R. L., 2001, \apj, 551, 852

\bibitem[Hibbard \& Mihos(1995)]{hm95} Hibbard, J. E., \& Mihos, J. C.,
1995, \aj, 110, 140

\bibitem[Hibbard et al.(2001)]{h01} Hibbard, J. E., van der Hulst, J. M.,
 Barnes, J. E., \& Rich, R. M., 2001, \aj, 122, 2969

\bibitem[Ho \& Filippenko(1996)]{hf96} Ho, L. C., \& Filippenko, A. V., 1996, 
  \apj, 466, L83

\bibitem[Holtzman et al.(1992)]{h92} Holtzman, J. A., et al., 1992, \aj, 103, 
  691

\bibitem[Hunter et~al.(1995)]{h95} Hunter, D. A., Shaya, E. J., 
Holtzman, J. A., Light, R. M., O'Neil, E. J., \& Lynds, R., 1995, 
\apj, 448, 179

\bibitem[Hunter et al.(2000)]{h00}Hunter, D. A., O'Connell, R. W., Gallagher, 
J. S., \& Smecker-Hane, T. A., 2000, \aj, 120, 2383

\bibitem[Johansson et al.(1998)]{j98}
Johansson, L. E. B., et al., 1998, \aap, 331, 857  

\bibitem[Johnstone et al.(2000)]{j00} Johnstone, D., Wilson, C. D.,
 Moriarty-Schieven, G., Joncas, G., Smith, G., Gregersen, E.,
\& Fich, M., 2000, \apj, 545, 327

\bibitem[Jog \& Solomon(1992)]{js92} Jog, C. \& Solomon, P. M., 1992, \apj, 
  387, 152

\bibitem[Kennicutt(1998)]{ken98}Kennicutt, R. C., 1998, \apj, 498, 541

\bibitem[Kramer et al.(1998)]{k98} Kramer, C., Stutzki, J., R\"ohrig, R., 
\& Corneiussen, U., 1998, \aap, 329, 249

\bibitem[Kutner et al.(1997)]{k97}
Kutner, M. L., et al., 1997, \aaps, 122, 255

\bibitem[Kwan(1979)]{k79}Kwan, J., 1979, \apj, 229, 567 

\bibitem[Lada et al.(1991a)]{l91a} Lada, E. A., Evans, N. J., Depoy, D. L., 
  \& Gatley, I., 1991, \apj, 371, 171

\bibitem[Lada et al.(1991b)Lada, Bally, \& Stark]{l91b}
Lada, E. A., Bally, J., \& Stark, A. A., 1991, \apj, 368, 432

\bibitem[Lamers et al.(2002)]{l02}Lamers, H. J. G. L. M., Panagia, N.,
Scuderi, S., Romaniello, M., Spaans, M., de Wit, W. J.,
\& Kirshner, R., 2002, \apj, 566, 818

\bibitem[Larsen(2000)]{l00} Larsen, S. S., 2000, \mnras, 319, 893

\bibitem[Larsen \& Richtler(2000)]{lr00} Larsen, S. S., \& Richtler, T.,
2000, \aap, 354, 836

\bibitem[Liang et al.(2001)]{l01} Liang, M. C., Geballe, T. R., Lo, K. Y.,
  \& Kim, D.-C., 2001, \apj, 549, L59

\bibitem[Massey \& Hunter(1998)]{mh98}
Massey, P., \& Hunter, D. A., 1998, \apj, 493, 180

\bibitem[McLaughlin \& Pudritz(1996)]{mp96}McLaughlin, D. E., \&
Pudritz, R. E., 1996, \apj, 457, 578

\bibitem[Mengel et al.(2002)]{m02}Mengel, S., Lehnert, M. D., Thatte, N.
\& Genzel, R., 2002, \aap, 383, 137

\bibitem[Mihos et al.(1993)Mihos, Bothun, \& Richstone]{m93}Mihos, J. C., 
Bothun, G. D., \& Richstone, D. O., 1993, \apj, 418, 82

\bibitem[Miller et al.(1997)]{m97}Miller, B. W., Whitmore, B. C., Schweizer, 
F., \& Fall, S. M., 1997, \aj, 114, 2381

\bibitem[Mirabel et al.(1998)]{m98}Mirabel, I. F. et al. 1998, \aap, 333, L1

\bibitem[Motte et al.(1998)]{mo98}Motte, F., Andr\'e, P., \& Neri, R. 1998, 
\aap, 336, 150

\bibitem[Motte et al.(2001)]{m01}Motte, F., Andr\'e, P., Ward-Thompson, D., 
\& Bontemps, S., 2001, 
\aap, 372, L41

\bibitem[Neff \& Ulvestad(2000)]{nu00}Neff, S. G. \& Ulvestad, J. S., 2000, 
\aj, 120, 670

\bibitem[O'Connell et al.(1995)]{o95} O'Connell, R. W., Gallagher, J. S.,
  Hunter, D. A., \& Colley, W. N., 1995, \apj, 446, L1

\bibitem[Planesas et al.(1991)]{p91}Planesas, P., Scoville, N., \& Myers,
S. T., 1991, \apj, 369, 364


\bibitem[Rand \& Kulkarni(1990)]{rk90}Rand, R. J., \& Kulkarni, S. R., 1990, 
  \apj, 349, L43

\bibitem[Rand(1993)]{r93}Rand, R. J., 1993, \apj, 410, 68

\bibitem[Rand(1995)]{r95}Rand, R. J., 1995, \aj, 109, 2444

\bibitem[Rand et al.(1999)]{r99}Rand, R. J., Lord, S. D., \& Higdon, J. L., 
1999, \apj, 513, 720

\bibitem[Rice et al.(1988)]{r88}Rice, W., Lonsdale, C. J., Soifer, B. T.,
Neugebauer, G., Koplan, E. L., Lloyd, L. A., de Jong, T.,
\& Habing, H. J., 1988, \apjs, 68, 91

\bibitem[Sakamoto(1996)]{sa96}Sakamoto, K., 1996, \apj, 471, 173

\bibitem[Sakamoto et al.(1999)]{s99}Sakamoto, K., Scoville, N. Z., Yun, M. S.,
Crosas, M., Genzel, R., \& Tacconi, L. J., 1999, \apj, 514, 68

\bibitem[Salpeter(1955)]{s55}Salpeter, E. E., 1955, \apj, 121, 161

\bibitem[Sanders \& Mirabel(1985)]{sm85}Sanders, D. B. \& Mirabel, I. F. 1985,
  \apj, 298, L31

\bibitem[Sanders, Scoville, \& Solomon(1985)]{sss85} Sanders, D. B., 
  Scoville, N. Z., \& Solomon, P. M., 1985, \apj, 289, 373

\bibitem[Sault, Teuben, \& Wright(1995)]{stm95} Sault,
R.~J., Teuben, P.~J., \& Wright, M.~C.~H.\ 1995, ASP Conf.~Ser.~77:
Astronomical Data Analysis Software and Systems IV, 4, 433

\bibitem[Schinnerer et al.(2000)]{s00}Schinnerer, E., 
Eckart, A., Tacconi, L. J., Genzel, R., \&  Downes, D.,
2000, \apj, 533, 850

\bibitem[Schweizer et al.(1996)]{s96} Schweizer, F., Miller, B.,
  Whitmore, B. C., \& Fall, S. M., 1996, \aj, 112, 1839

\bibitem[Scoville et al.(1993)]{s93} Scoville, N. Z., Carlstrom, J. E.,
Chandler, C. J., Phillips, J. A., Scott, S. L.,
 Tilanus, R. P. J., \& Wang, Z., 1993, \pasp, 105, 1482

\bibitem[Scoville et al.(2001)]{s01} Scoville, N. Z., Polletta, M.,
Ewald, S., Stolovy, S. R., Thompson, R., \& Rieke, M., 2001, \aj, 122, 3017

\bibitem[Solomon et al.(1987)]{s87}Solomon, P. M., Rivolo, A. R.,
Barrett, J., \& Yahil, A. 1987, \apj, 319, 730

\bibitem[Solomon et al.(1997)]{sdrb97}Solomon, P. M., Downes, D., Radford, S. 
J. E., \& Barrett, J. W., 1997, \apj, 478, 144

\bibitem[Smith \& Gallagher(2001)]{sg01}Smith, L. J, \& Gallagher, J. S.,
2001, \mnras, 326, 1027

\bibitem[Stanford et al.(1990)]{s90}Stanford, S. A., Sargent, A. I., Sanders, 
  D. B. \& Scoville, N. Z. 1990, \apj, 349, 492

\bibitem[Strong et al.(1988)]{s88}Strong, A. W. et al. 1988, \aap, 207, 1

\bibitem[Testi \& Sargent(1998)]{ts98} Testi, L., \& Sargent, A., 1998, \apj, 
508, L91

\bibitem[Thornley \& Mundy(1997)]{t97}Thornley, M. D., \& Mundy, L. G., 1997, 
\apj, 484, 202

\bibitem[Tully \& Fisher(1988)]{tf88}Tully, R. B., \& Fisher, J. R., 1988,
``Catalog of Nearby Galaxies'', Cambridge University Press
\apj, 484, 202

\bibitem[Turner et al.(1998)]{t98} Turner, A., et al., 1998, \apj, 505, 207

\bibitem[Vigroux et al.(1996)]{v96}Vigroux, L. et al. 1996, \aap, 315, L93

\bibitem[Whitmore(2002)]{w02} Whitmore, B. C., 2002, in ``Extragalactic
Star Clusters'', eds. D. Geisler, E. K. Grebel, \& D. Minniti,
[San Francisco: Astronomical Society of the Pacific], 367.

\bibitem[Whitmore(2003)]{w01} Whitmore, B. C., 2003, in ``A Decade of Hubble 
Space Telescope Science'' 
eds. M. Livio, K. Noll, M. Stiavelli, [Cambridge University Press], 
in press (astro-ph/0012546)

\bibitem[Whitmore et al.(1993)]{w93} Whitmore, B. C., Schweizer, F., Leitherer,
  C., Borne, K., \& Robert, C. 1993, \aj, 106, 1354

\bibitem[Whitmore \& Schweizer(1995)]{ws95} Whitmore, B. C. \& Schweizer, F., 
  1995, \aj, 109, 960

\bibitem[Whitmore et al.(1999)]{w99} Whitmore, B. C., Zhang, Q.,
  Leitherer, C., Fall, S. M., Schweizer, F., \& Miller, B. W., 1999, 
  \aj, 118, 1551

\bibitem[Whitmore \& Zhang(2002)]{wz02} Whitmore, B. C. \& Zhang, Q., 2002,
\aj, 124, 1418

\bibitem[Williams et al(1994)Williams, de Geus, \& Blitz]{w94}Williams, J. P., de Geus, E. J. \& Blitz, L. 1994, \apj, 428, 693

\bibitem[Wilson(1995)]{w95}
Wilson, C. D., 1995, \apj, 448, L97

\bibitem[Wilson \& Scoville(1990)]{ws90} Wilson, C. D. \& Scoville, N. Z.,
1990, \apj, 363, 435


\bibitem[Paper I()]{w00} Wilson, C. D., Scoville, N., Madden, S. C.,
  \& Charmandaris, V., 2000, \apj, 542, 120 (Paper~I)

\bibitem[Zhang \& Fall(1999)]{zf99}  Zhang, Q., \& Fall, S. M., 1999, 
  \apj, 527, L81

\bibitem[Zhang et al.(2001)Zhang, Fall, \& Whitmore]{z01}
Zhang, Q., Fall, S. M., \& Whitmore, B. C., 2001, \apj, 561, 727

\end{thebibliography}
\end{document}